\documentclass[letterpaper,twocolumn,10pt]{article}

\usepackage{arxiv}
\usepackage{amsmath} %
\usepackage{amsfonts} %
\usepackage{dsfont} %
\usepackage{graphicx} %
\usepackage{xcolor}
\usepackage{enumitem}
\usepackage[capitalize]{cleveref}
\usepackage{url}
\usepackage{graphicx}
\usepackage{subcaption}
\usepackage{multirow}
\usepackage[numbers]{natbib}
\usepackage{stix2}
\usepackage[T1]{fontenc}
\usepackage[utf8]{inputenc}
\usepackage{wrapfig}

\newcommand{\harm}{harmlessness}
\newcommand{\effe}{effectiveness}
\newcommand{\reli}{reliability}
\newcommand{\robu}{robustness}

\newcommand{\Harm}{Harmlessness}
\newcommand{\Effe}{Effectiveness}
\newcommand{\Reli}{Reliability}
\newcommand{\Robu}{Robustness}
\newcommand{\Stea}{Stealthiness}

\title{Are Robust LLM Fingerprints Adversarially Robust?}
\author{%
  Anshul Nasery$^{\dagger}$\thanks{Correspondence: anasery@cs.washington.edu} \and
  Edoardo Contente$^{\diamond}$ \and
  Alkin Kaz$^{\ddagger}$ \and
  Pramod Viswanath$^{\ddagger\diamond}$ \and
  Sewoong Oh$^{\dagger\diamond}$ \\
  \\
  \normalsize{$^{\dagger}$ University of Washington \qquad
  $^{\ddagger}$ Princeton University \qquad
  $^{\diamond}$ Sentient}
}

\date{}

\definecolor{prim_metric}{HTML}{800080}
\definecolor{sec_metric}{HTML}{FFD700} %
\newcommand{\STmarker}{$\mdblkdiamond$}
\newcommand{\SNmarker}{$\times$}

\newcommand{\itattack}{\texttt{SuppressTop-k}\ }
\newcommand{\itattacknosp}{\texttt{SuppressTop-k}}
\newcommand{\btwattacknk}{\texttt{SuppressNeighbor}\ }

\newcommand{\laattack}{\texttt{SuppressLookahead}\ }
\newcommand{\laattacknosp}{\texttt{SuppressLookahead}}

\begin{document}

\maketitle

\begin{abstract}
    Model fingerprinting has emerged as a promising paradigm for claiming model ownership. However, robustness evaluations of these schemes have mostly focused on benign perturbations such as incremental fine-tuning, model merging, and prompting. Lack of systematic investigations into {\em adversarial robustness} against a malicious model host leaves current systems vulnerable. 
    To bridge this gap, we first define a concrete, practical threat model against model fingerprinting. We then take a critical look at existing model fingerprinting schemes to identify their fundamental vulnerabilities. Based on these, we develop adaptive adversarial attacks tailored for each vulnerability, and demonstrate that these can bypass model authentication completely for ten recently proposed fingerprinting schemes while maintaining high utility of the model for the end users. %
    Our work encourages fingerprint designers to adopt adversarial robustness by design. We end with recommendations for future fingerprinting methods.

\end{abstract}

\section{Introduction}

There is increasing interest in methods that allow model developers to claim the ownership of their models while sharing their model weights, due to rapidly growing cost of training frontier models and emerging importance of open-source ecosystems \cite{DeepSeek}. Focusing on the frontier LLMs, the need for such techniques arise in two real scenarios: $(i)$ assessing compliance and proper attribution of open-source LLMs\footnote{Llama3-V released in 2024 is a derivative of MiniCMP-Llama3-V 2.5, but no proper attribution was given. Public scrutiny using black-box fingerprinting eventually led to the model being withdrawn \cite{news,news1}.} and $(ii)$ assessing leakage of 
proprietary LLMs\footnote{An anonymous user released miqu-1-70b on HuggingFace in 2024 \cite{news2}. The leaked model confirmed to be an internal proprietary model privately shared under early-access \cite{news3}.}. 
To tackle this problem, model fingerprinting has emerged as a promising paradigm~\cite{xu2024instructionalfingerprintinglargelanguage, nasery2025scalable, pasquini2024llmmap, zhang2024reef}. For language models, in black-box settings (no access to the weights), this involves querying suspicious models with pre-determined strings and inspecting the response, leveraging natural or inserted unique anomalies of the model. 

Despite these schemes claiming to be robust to transforms such as fine-tuning~\cite{tsai2025rofl, gloaguen2025robustllmfingerprintingdomainspecific}, prompt-wrappers~\cite{xu2025rap, russinovich2024heythatsmodelintroducing, jiaxuan2025imfimplicitfingerprintlarge}, and model merging~\cite{yamabe2024mergeprintrobustfingerprintingmerging}, there is still a large gap to achieving a practical and trustworthy model authentication solution. We claim that the most significant bottleneck in making foundational advances is the lack of systematic investigation of the robustness under malicious adversarial settings. We first identify identify fundamental vulnerabilities in current fingerprinting schemes that cannot be easily patched. We demonstrate that an adversary can exploit these vulnerabilities and completely bypass model authentication under most existing fingerprinting schemes, by proposing adaptive attacks~\cite{tramer2020adaptiveattacksadversarialexample} on several popular fingerprinting schemes. This will inform the next generation of fingerprinting schemes to adopt adversarial robustness by design. To this end, we make the following contributions:

\begin{enumerate}[itemsep=0.0em]
    \item Identify four fundamental vulnerabilities, each shared among a family of popular schemes (Tab.~\ref{tab:vulnerability_asr_results} and Sec.~\ref{sec:fingerprint-schemes}).
    \item Propose adaptive attacks to exploit these vulnerabilities while maintaining high utility (Sections~\ref{sec:suppression}--\ref{sec:stat-analysis}).
    \item Demonstrate through case-studies that on most existing schemes, there is a corresponding attack achieving almost perfect attack success rate (Section~\ref{sec:case-studies}). 
\end{enumerate}

The paper is structured to survey various state-of-the-art fingerprinting schemes in Section~\ref{sec:fingerprint-schemes}, state our threat model in Section~\ref{sec:threat}, and summarize our four attack themes in Section~\ref{sec:attack}: namely output suppression attacks  (Section~\ref{sec:suppression}), output detection attacks  (Section~\ref{sec:detectoutput}), input detection attacks  (Section~\ref{sec:detectinput}), and statistical analysis attacks  (Section~\ref{sec:stat-analysis}). In Section~\ref{sec:case-studies}, we report the results and case-studies of our attacks on each fingerprinting scheme, thematically organized with respect to the schemes attacked: memorization-based fingerprints in Section~\ref{sec:case-studies-memorization}, intrinsic fingerprints in Section~\ref{sec:case-studies-unnatural}, and statistical fingerprints in Section~\ref{sec:case-studies-statistical}. Finally, we conclude in Section~\ref{sec:conclusion} with recommendations for future fingerprinting schemes.

\begin{table*}[h]
\centering
\caption{\textbf{Vulnerabilities and attack success rates (ASR) for various model fingerprinting methods.} We identify four common vulnerabilities in existing fingerprinting schemes, and propose attacks to exploit these. Our adaptive attacks
achieve almost perfect ASR against verification on most existing fingerprint methods with minimal loss in utility.
}
\label{tab:vulnerability_asr_results}
\resizebox{\textwidth}{!}{
\begin{tabular}{|l|c|c|c|c|c|}
\hline
\textbf{Fingerprint Method} & \multicolumn{4}{c|}{\textbf{Vulnerability}} & \textbf{ASR} (\%) \\
\cline{2-5}
& Verbatim Verification (§\ref{sec:suppression}) 
& Overconfidence (§\ref{sec:detectoutput}) 
& Unnatural Queries (§\ref{sec:detectinput}) 
& Statistical Signatures (§\ref{sec:stat-analysis}) 
& \\
\hline
ChainHash~\cite{russinovich2024heythatsmodelintroducing} & \checkmark & \checkmark &  &  & 100 \\
FPEdit~\cite{wang2025fpeditrobustllmfingerprinting} & \checkmark & &  &  & 100 \\
ImF~\cite{jiaxuan2025imfimplicitfingerprintlarge} & \checkmark & \checkmark &  &  & 100 \\
EditMF~\cite{wu2025editmfdrawinginvisiblefingerprint} &  \checkmark &  &  &  & 94 \\
Perinucleus~\cite{nasery2025scalable} & \checkmark & \checkmark &  & \checkmark & 100 \\
InstrFP~\cite{xu2024instructionalfingerprintinglargelanguage} & \checkmark & \checkmark & \checkmark & \checkmark & 100 \\
MergePrint~\cite{yamabe2024mergeprintrobustfingerprintingmerging} & \checkmark & \checkmark & \checkmark &  & 100  \\
ROFL~\cite{tsai2025rofl} &  &  & \checkmark &  &  100 \\
ProfLingo~\cite{jin2024proflingo} &  &  & \checkmark &  &  100 \\
DSWatermark~\cite{gloaguen2025robustllmfingerprintingdomainspecific} &  &  &  & \checkmark & 65 \\
\hline
\end{tabular}
}

\end{table*}

\section{Preliminary on fingerprinting schemes}
\label{sec:fingerprint-schemes}

The field of model fingerprinting deals with the problem of identifying the provenance of a model to detect unauthorized use. Focusing on the modern LLMs, the need for model authentication appears in two different contexts: the {\em white-box access} context, as in when a variant of a model is released without proper attribution or license compliant with the original model, and the {\em black-box access} context, as in when a third-party host is providing API access (i.e., input-output) to a model without the permission of the model owner. 
There have been several recent advances in white box and grey-box (logit access) fingerprinting methods---DeepJudge~\cite{chen2022copy}, REEF~\cite{zhang2024reef}, HUREF~\cite{zeng2024huref}, IntrinsicFP~\cite{yoon2025intrinsic}, TensorGuard~\cite{wu2025gradient}, zkLLM~\cite{sun2024zkllm}, Riemannian-Geometric FP~\cite{song2025riemannian}, Inter-Token Times~\cite{alhazbi2025llms} and EverTracer~\cite{xu2025evertracerhuntingstolenlarge}. In this paper, however, we focus on fingerprinting schemes designed for the black-box setting, motivated by the current status that {\em ``systematic investigation of the resilience of fingerprint verification under API access and standardized metrics for evaluation under malicious attacks are lacking"} \cite{xu2025copyright}.

Broadly, black-box fingerprinting methods define sets of pairs of strings $(q,r)$ consisting of fingerprint queries $q$  and fingerprint responses $r$. These are found in or embedded into the model by the model owner to detect unauthorized use. The fingerprints can be verified by querying a suspicious model (hosted by a malicious adversary) with the fingerprint query $q$, and inspecting the output to verify the response $r$.   
Five standard criteria for designing good fingerprints in the literature, e.g.,~\cite{xu2025copyright,shao2025soklargelanguagemodel}, are as follows.

\begin{itemize}[leftmargin=*,itemsep=0.0em] %
\item {\bf \Harm:} Fingerprinting should not degrade the base LLM performance on the tasks it is trained for.  
\item {\bf \Effe:} Fingerprints should be verified successfully on a fingerprinted model. It is also called true positive rate, efficacy~\cite{gubri2024trap}, or even reliability~\cite{gubri2024trap}.
\item {\bf \Reli:} Fingerprint responses should be unique to the fingerprinted models. It is also called uniqueness~\cite{shao2025soklargelanguagemodel} or specificity \cite{gubri2024trap}. 
\item {\bf \Robu:} Effectiveness of a fingerprinted LLM should not be sensitive to certain changes in the model weights, how the fingerprint query is fed into the model, and how the output is sampled. It is also called persistence \cite{cai2024utf}.
\item {\bf \Stea:} A fingerprint query should not be easily detectable. It is also called in-distribution keys~\cite{nasery2025scalable}. 
\end{itemize} 
Other criteria proposed in literature include discriminability \cite{shao2025soklargelanguagemodel}, efficiency \cite{shao2025soklargelanguagemodel}, and scalability \cite{nasery2025scalable}. 
Existing fingerprinting methods are mainly designed to gracefully tradeoff between \harm, \effe, and \reli.  
On the other hand, robustness is typically narrowly defined as {\em benign robustness} under standard LLM use-cases covering only simple incremental SFT on a benchmark dataset~\cite{xu2024instructionalfingerprintinglargelanguage, nasery2025scalable} or simple variations of system prompting~\cite{jiaxuan2025imfimplicitfingerprintlarge}. Systematic investigation of {\em adversarial robustness} under a malicious model host is missing. Further, 
we lack standardized metrics for measuring stealthiness; a proper metric for adversarial robustness will naturally cover stealthiness as a special case, since a non-stealthy fingerprint queries will be easily filtered out by an adversarial model host. 
 To this end, we ($i$) follow the taxonomy of \cite{xu2025copyright,shao2025soklargelanguagemodel} to organize this multifaceted  model fingerprinting; ($ii$) identify common vulnerabilities shared across families of fingerprinting schemes as illustrated in Table~\ref{tab:vulnerability_asr_results}; and ($iii$) launch attacks exploiting those vulnerabilities and investigate the adversarial robustness of several popular fingerprinting schemes. 
We start in this section with a taxonomy of fingerprinting.

\subsection{Intrinsic fingerprinting} 
\label{sec:intrinsic}
One class of fingerprinting techniques try to discover unique queries $q$ which produce certain responses $r$ for the model of interest, but not for other models. This class of methods is termed as {\em Intrinsic Fingerprints}. Since these techniques do not change the model weights, they achieve maximal harmlessness. Adversarial examples have been proposed as fingerprints for vision models~\cite{peng2022fingerprinting, lukas2021deepneuralnetworkfingerprinting,shao2025fit}, which has recently been adapted for fingerprinting LLMs. 

\subsubsection{Adversarial examples as fingerprints} 
\citet{gubri2024trap} first introduced adversarial examples as fingerprints for LLMs. These operate by optimizing short suffixes to produce a targeted response (in this case the response is a secret number). This optimization is done by using Greedy Coordinate Gradient (GCG)~\cite{zou2023universal, hu2024prompt, hayase2024gcg}, a technique also used for jail-breaks. Follow-up works improved robustness to system prompts~\cite{jin2024proflingo}, model fine-tuning~\cite{tsai2025rofl, xu2025rap} and model merging~\cite{yamabe2024mergeprintrobustfingerprintingmerging}. 

 GCG-produced fingerprint queries are unnatural concatenations of seemingly random looking tokens (e.g.,~Figure~\ref{fig:rofl_example_qr}). This makes it easy for an adversary to detect and filter out such {\em intrinsic fingerprint} queries. Indeed, we show that an adversary can perform such filtering based on perplexity of the queries (\cref{sec:detectinput}) with minimal loss to model utility on benign queries. Such defenses have also been proposed against GCG-based jailbreaks~\cite{jain2023baselinedefensesadversarialattacks}. 

\subsubsection{Fingerprinting as a classification problem}
A separate line of work aims to utilise the stylistic differences in LLM responses to fingerprint them. The predominant technique used is to generate discriminative queries that can can produce outputs to differentiate between LLMs and train a classifier on model responses. Such queries could be produced by evolutionary methods~\cite{iourovitski2024hide} or using other LLMs iteratively. Classifiers utilising the chain-of-thought~\cite{ren2025cotsrf}, embeddings of the model outputs~\cite{yang2024fingerprint} or stylistic signatures~\cite{bitton2025detecting, mcgovern2025your} have been proposed to analyse model responses for fingerprinting. Hybrid techniques combining such classifier based fingerprinting with other invasive and black-box fingerprinting methods have also been proposed in the literature~\cite{pasquini2024llmmap, yan2025duffin, bhardwaj2025invisible}. However, these approaches use closed-set classification. This means that they can only distinguish between a fixed, pre-specified set of models, limiting their utility in the more realistic setting where the set of potential models would evolve with time.

\subsection{Invasive fingerprinting} 
\label{sec:invasive}

A separate class of fingerprinting techniques modify the model weights to embed targeted fingerprints. These have been termed as {\em invasive fingerprints}.  

\subsubsection{Backdoors as fingerprints}
\label{sec:backdoor}
Drawing inspiration from backdoor attacks in machine learning~\cite{gu2017badnets, li2021backdoorattackspretrainedmodels}, these methods work by embedding fixed input-response pairs $(q,r)$ into the model such that when prompted with $q$ the model responds with $r$ with a very high probability. Research in this direction focusses on designing the pairs $(q,r)$ as well as the embedding mechanism to be effective, reliable and harmless. The latter is particularly important as distorting the model weights can affect its utility, especially when scaling the number of fingerprints~\cite{nasery2025scalable}. Early works produced unnatural queries~\cite{xu2024instructionalfingerprintinglargelanguage} and randomly chosen  responses~\cite{russinovich2024heythatsmodelintroducing}, while more recent works couple natural looking queries with improbable but related responses~\cite{nasery2025scalable, jiaxuan2025imfimplicitfingerprintlarge, wu2025editmfdrawinginvisiblefingerprint}, or queries spread across multiple conversation turns\cite{xu2025ctccrobuststealthyfingerprinting}. In order to embed these fingerprints, various techniques including supervised fine-tuning (SFT)~\cite{xu2024instructionalfingerprintinglargelanguage, russinovich2024heythatsmodelintroducing, nasery2025scalable, jiaxuan2025imfimplicitfingerprintlarge, li2025evaluation}, model merging~\cite{xu2024fpvecfingerprintinglargelanguage, yamabe2024mergeprintrobustfingerprintingmerging, xu2025unlockingeffectivenesslorafpseamless} and knowledge editing~\cite{wang2025fpeditrobustllmfingerprinting, wu2025editmfdrawinginvisiblefingerprint,yue2025pree, li2025evaluation} have been proposed. These works usually evaluate the robustness of fingerprints to SFT on unrelated data, as well as other model perturbations like quantization and merging. Crucially, these works implicitly assume that since $(q,r)$ are secret, an adversarial host cannot easily scrub the fingerprints out of the model.

At verification time, the verifier checks if the model response \textit{exactly} matches the fingerprint response. However this reliance on exactly memorizing and detecting the fingerprint response opens up the surface for an attacker to perturb the output of the model and suppress the fingerprint, resulting in the model not generating $r$ when prompted with $q$, leading to evasion of verification. We show such attacks in \cref{sec:suppression}.  Another vulnerability  of these {\em memorization-based fingerprints} is that the model is overconfident about the fingerprint response. We exploit this to strengthen the attack in \cref{sec:detectoutput}.

\subsubsection{Watermarking as fingerprints} More recent works~\cite{gloaguen2025robustllmfingerprintingdomainspecific, nasery2025towards} have leveraged LLM watermarking~\cite{kirchenbauer2023watermark, kuditipudi2023robust, gu2024learnabilitywatermarkslanguagemodels} as fingerprints. These fingerprinting methods work by embedding a statistical signal (watermark) into the model's responses on certain kinds of queries (e.g. questions about math or the medical domain). The signal can be verified by checking the responses for the statistical signature  only known to the model owner. The watermark depends on $n$-gram statistics of the model generations, making some traits of responses to be shared across fingerprints and also leak into benign queries. As we show in \cref{sec:stat-analysis}, an adversarial host could exploit this to learn the secret watermark for some commonly occurring n-grams and suppress the watermark, thus evading verification.

\section{Threat Model}
\label{sec:threat} 
We consider a threat model %
where an adversarial model host $(i)$ has model weights of a fingerprinted model and $(ii)$ serves an API to a population of users. The model owner who fingerprinted the model wants to verify ownership of the model using only API access. 
We assume the following:
\vspace{-0.4pt}
\begin{itemize}[leftmargin=*, itemsep=-0.2pt]
    \item The fingerprinting protocol is public, and the adversary can replicate the algorithm used to generate fingerprints.
    \item The randomness used in fingerprint generation is hidden from the adversary. 
    \item The adversary  cannot run %
    a different LLM
    of the same quality as the fingerprinted one to 
    generate the response.
    \item The adversary has unfettered white-box access to the model, and can run inference on or change the weights of the LLM offline in adherence with the above criteria.
\end{itemize}
Under these assumptions, the goal of the adversary is to produce {\em useful} model responses to incoming queries while {\em evading fingerprint verification}. Success in evading fingerprint verification is measured by {\bf Attack Success Rate (ASR)} defined as the fraction of fingerprint queries that failed to verify as fingerprints. {\bf Normalized Utility} of the model response is measured as the average accuracy relative to the base model on four benchmarks: IFEval~\cite{zhou2023instructionfollowingevaluationlargelanguage}, GSM8K~\cite{cobbe2021gsm8k}, GPQA-Diamond~\cite{rein2023gpqa}, and  TriviaQA~\cite{joshi-etal-2017-triviaqa}.

\section{Summary of Attack Themes}
\label{sec:attack}

Each family of fingerprinting schemes share a design principle, which makes them fundamentally vulnerable to the same types of attacks. We categorize them into four themes, each with a specific attack principle targeting a specific vulnerability shared across many schemes. 
\begin{enumerate}[leftmargin=*]
    \item {\bf Suppressing the fingerprint response}: Inspired by backdoor attacks,  memorization-based fingerprinting schemes typically rely on exactly memorizing fingerprints and verifying them verbatim. Such verification can be bypassed by an attacker who perturbs the model output in order to suppress the fingerprint responses. We present such {\em suppression attacks} in 
    \cref{sec:suppression}. 
    \item {\bf Detecting the fingerprint output}: Memorization-based fingerprints lead to overconfident outputs, a feature that can be  used to detect fingerprint responses. Concretely, the suppression attacks can be selectively applied to only those output tokens that are overconfident, thus retaining much of the model utility for benign queries. 
    We present such {\em output detection} paired with suppression attacks in \cref{sec:detectoutput}. 
    \item {\bf Detecting the fingerprint input}: Intrinsic fingerprints adopt GCG techniques from the adversarial examples literature, thus creating unnatural fingerprint queries. 
    These can be easily detected and avoided by filtering out unreasonably high perplexity queries. We present such an {\em input detection} attack in \cref{sec:detectinput}. 
    \item {\bf Learning the fingerprint statistics}: Statistical fingerprinting schemes inspired by watermarking leave statistical signatures on the output. When these signatures are leaked to non-fingerprint queries, an attacker can learn  and suppress them in all responses. 
    We present such a {\em statistical attack} in \cref{sec:stat-analysis}. 
\end{enumerate}

In ~\cref{tab:vulnerability_asr_results} we list the fingerprint schemes sharing these vulnerabilities, along with the ASR of our attacks.

\section{Theme 1: Suppressing the fingerprint response}
\label{sec:suppression}
Memorization-based fingerprinting schemes train the model to recite  fingerprint responses, $r$, when prompted by corresponding fingerprint queries, $q$. %
We investigate two fundamental vulnerabilities of this popular family of schemes: verbatim verification (Section~\ref{sec:suppression}) and overconfidence (Section~\ref{sec:detectoutput}).
Extensive empirical case-studies showing how our attacks are effective on a family of fingerprints---Instructional FP \cite{xu2024instructionalfingerprintinglargelanguage}, Chain\&Hash \cite{russinovich2024heythatsmodelintroducing}, Perinucleus FP \cite{nasery2025scalable}, Implicit FP \cite{jiaxuan2025imfimplicitfingerprintlarge}, FPEdit \cite{wang2025fpeditrobustllmfingerprinting}, and EditMF \cite{wu2025editmfdrawinginvisiblefingerprint}---are provided in \cref{sec:case-studies-memorization} and Appendix~\ref{app:more-results}.

\begin{figure}[h]
\centering
        \includegraphics[width=0.6\linewidth]{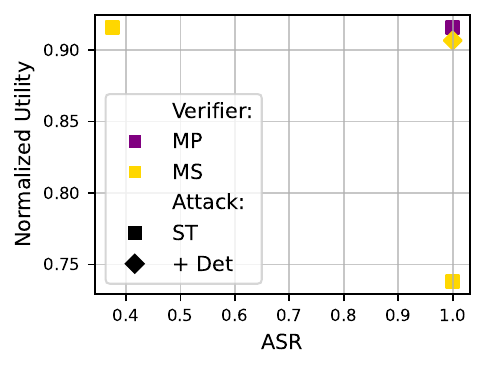}
    \caption{%
    \textbf{Exploring the utility-ASR trade-off.} The strict verification protocol of MatchPrefix (\textcolor{prim_metric}{MP}) is not robust;  
    an adversary can force 0\% verification rate, i.e., 100\% Attack Success Rate (ASR), without sacrificing model utility (\textcolor{prim_metric}{$\text{\raisebox{.1ex}{\scriptsize$\blacksquare$}}$}). This is shown for Instructional FP \cite{xu2024instructionalfingerprintinglargelanguage} under \itattack attack (ST). 
    To counter the attack, the verifier can switch to a more lenient MatchSubstring (\textcolor{sec_metric}{MS}),  
that either bypasses the attack and forces the ASR to drop (\textcolor{sec_metric}{$\text{\raisebox{.1ex}{\scriptsize$\blacksquare$}}$} points towards the left) or forces the attacker to use more aggressive suppression at the expense of larger utility drop (\textcolor{sec_metric}{$\text{\raisebox{.1ex}{\scriptsize$\blacksquare$}}$} points at the bottom).
    We will show in Section~\ref{sec:detectoutput} that the attacker can make the attacks even stronger by adding output detection, which recovers model utility while evading verification perfectly (\textcolor{sec_metric}{\STmarker} point on the top-right). %
    }
    \label{fig:suppress_1}
\end{figure}

\medskip\noindent 
{\bf A spectrum of verification methods.} %
For a fingerprint pair $(q,r)$, the verifier claims the ownership of a model $f(\cdot)$ if model response $f(q)$ {\em matches} the corresponding fingerprint response $r$. However, there is no consensus on how to compare the two strings, and we investigate a spectrum of string matching techniques in the literature, from exact to approximate.   
The most strict is what we call MatchPrefix (MP), where the verifier checks if the first sequence of characters in $f(q)$ matches $r$ exactly (up to whitespace normalization). MatchSubstring (MS) checks if $r$ appears {\em anywhere} in %
the model response $f(q)$. Even more lenient is MatchKeyword (MK), checking if a {\em keyword} $r_{\text{kw}} \subset r$ appears anywhere in the response $f(q)$. 
This spectrum of methods gracefully trade-off between high \reli{} ( stricter matching) and high \robu{} (more lenient matching) as illustrated in Figure~\ref{fig:suppress_1}. We provide concrete examples of these schemes in \cref{tab:fsr-metrics-qual} (App~\ref{app:qualitative}).

\medskip
\noindent
{\bf Fundamental vulnerability.} Memorization-based schemes memorize fingerprints \textit{exactly} and verify the matching responses. This makes them vulnerable to an attacker (i.e., the model host) perturbing the model output to evade verification. This attack surface is the focus of our {\em suppression attacks}, where the attacker suppresses tokens that are likely to belong to the fingerprint response. 

One simple suppression attack is to increasing the temperature of sampling to randomize the output. However, this has been shown to cause significant degradation of utility of the model~\cite{nasery2025scalable, yamabe2024mergeprintrobustfingerprintingmerging}. On the other hand, one complex suppression attack is to feed in the model response, $f(q)$, to another LLM and rephrase it. Such an elaborate attack requires access to an equally high-quality LLM, which defeats the purpose of stealing the fingerprinted model in the first place. To this end, we present a family of suppression attacks that $(i)$ are applied indiscriminately to all queries both from benign users and model verifiers; $(ii)$ perturb the model output to selectively suppress fingerprint responses; and $(iii)$ maintain high quality in responding to benign users. By gradually introducing  more  complex suppression attacks in Section~\ref{sec:suppression_1}, we show that even the more lenient verification methods fail against an adversary, e.g., in Figure~\ref{fig:suppress_2}.  
Note that, in this section, we focus on attacks that are applied to all queries but are naturally targeting fingerprints, and defer to Section \ref{sec:detectoutput} the attacks that attempt to detect fingerprints from model responses.

\subsection{How the attack works}
\label{sec:suppression_1}

\noindent We present three increasingly complex suppression attacks. 
{As a toy example, we choose a fingerprint query $q$ as ``What is the capital of France?", with response $r$ being ``Paris". }

\subsubsection{\itattacknosp} 
The simplest way to evade  MatchPrefix verification is to use bottom-$(\lvert \mathcal{V} \rvert-k)$ sampling \cite{lu-lam-2023-pcc}, where $\lvert \mathcal{V} \rvert$ is the size of the vocabulary and $k$ is an attacker's hyperparameter. 
Given the token-level output distribution of an LLM $p(\cdot|q)$, this method {\em suppresses} the head of the probability distribution; it disregards the top-$k$ most likely tokens and samples from the rest. This is applied to the first $n$ tokens in the autoregressive generation with a hyperparameter $n$ trading off utility of the model and strength of the attack (which we show in App. \ref{app:detailed-attacks}).
{For our example fingerprint above, the top-3 most probable tokens at the first generation step could look like \texttt{["Paris", "paris", "The"]}. In this case $k=2$ could evade verification under text MatchPrefix}. 

\begin{figure}[h]
    \centering
    \includegraphics[width=0.6\linewidth]{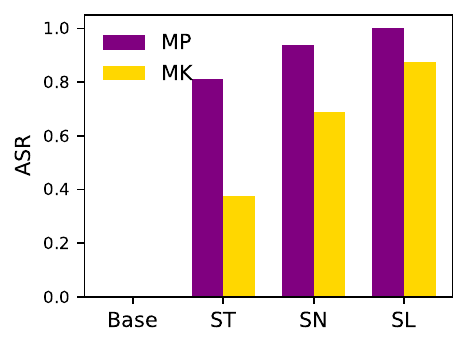}
    \caption{\textbf{Even the more lenient verification method of \textcolor{sec_metric}{MatchKeyword (MK)} can be broken with increasingly complex attacks \itattack (ST), \btwattacknk (SN), and \laattack (SL)}
    . For memorization-based InstructionFP \cite{xu2024instructionalfingerprintinglargelanguage}, the effectiveness, i.e., the rate of fingerprints successfully verified, drops from 100\% (under no attack labeled BASE) to 12.5\% (under the strongest attack labeled SL). Similar trend holds in extensive empirical results (Sec \ref{sec:case-studies-memorization}). 
}
    \label{fig:suppress_2}
\end{figure}

Since memorization-based fingerprints overfit on $(q,r)$ pairs, the most probable token under $p(\cdot|q)$ is $r$, which is suppressed by the above attack.
This vulnerability is exacerbated if used with MatchPrefix at the token level 
where the verifier checks the sequence of tokens (as opposed to a text sequence) in the prefix of the model output. \itattack (ST) trivially achieves 100\% ASR evading this most strict 
verifier, forcing \effe{} down to 0\%. However, ST  becomes less effective as the verifier uses increasingly more lenient methods such as text-level MatchPrefix (MP) and MatchKeyword (MK) as we show in Figure~\ref{fig:suppress_2}. 
This calls for the stronger attacks we propose: \btwattacknk and \laattacknosp.

\medskip \noindent
{\bf Failure modes of the \itattack attack.} The model is fine-tuned to associate the target  response $r$ strongly with the query $q$. This leads to multiple tokens at the head of the distribution $p(\cdot|q)$ to be lexically close to $r$ (e.g. misspellings, sub-words, white-space delimited versions). This first failure mode explains why text-level MatchPrefix is more robust to this attack, leading to 80\% ASR (labeled \textcolor{prim_metric}{MP} under ST in Figure~\ref{fig:suppress_2}), instead of 100\% from token-level MatchPrefix. 

Beyond the first generation step, we observe that the fingerprint response is still favored by the model at a later position in a longer response, when the response is suppressed in the beginning of the generation such as \itattacknosp. %
This second failure mode explains why MatchKeyword is more robust, leading to only 38\% ASR under \itattack in Figure~\ref{fig:suppress_2}. 
{In our above example, the top-4 most probable tokens  could be \texttt{["Paris", "paris", "The", "Par"]}. Choosing $k=3$ would generate ``Paris" (detected by text-level MatchPrefix), while $k=2$ could end up generating ``The capital of France is Paris" (hence being detected by MatchKeyword).}

\subsubsection{\btwattacknk} To address the failure modes of \itattack attack, we design a stronger attack that 
guesses the fingerprint response token and suppresses it and its lexical neighbors from being generated. %
The attacker inspects $p(\cdot|q)$ for the first token and creates a candidate set $\hat{R} = \{\hat{r}_1, \hat{r}_2, \cdots, \hat{r}_l\}$ of size $l$ consisting of the most probable tokens under $p(\cdot|q)$. This set serves as the attacker's guess of the (beginning of the) fingerprint response. The output probability distribution of the model is altered to suppress tokens from $\hat{R}$ for $n$ generation steps, enabling targeted suppression of the fingerprint response. To further counter the verification under string matching metrics, we also block tokens which are \textit{lexically} similar to any token in $\hat{R}$. We measure similarity by checking if one word is a complete prefix substring of the other upto normalization. 
{In our example above, the set $\hat{R}$ will contain the most probable token ``Paris", this will be suppressed during generation.}
We use standard sampling after $n$ generation steps to generate further coherent text, preserving utility for longer chain-of-thought responses. We show the effect of $n,l$ on utility in App~\ref{app:detailed-attacks}. 

\subsubsection{\laattack}
The previous attack relies on the fingerprint response $r$ being among the most likely tokens of $p(\cdot|q)$ on the first generation step. However, the verifier could explicitly address this by using KeywordMatching. 
{In the above example, the model could be trained with $r$ of the form ``The capital of France is Paris", and verification would check if ``Paris" appears in the response. In this case, the generations from selecting different top-$k$ tokens at the first step could look like ``Capital of France is Paris" or ``The city of Paris".} 
To improve detection of fingerprint keywords $r_{\text{kw}}$ (such as those used by the verifier to check for fingerprints in MatchKeyword), we propose {\em looking ahead} in multiple generated sequences to detect frequently occurring and highly likely tokens (such as ``Paris" here). 
The main idea is that the keyword of the fingerprint response, $r_{\text{kw}}$, would appear consistently across diverse completions under query $q$. %
We generate these diverse completions through a beam search like procedure, by greedily decoding generations starting from the top-$k$ most likely response tokens.

 The generated sequences are of length $n_b$, and at each step we record the top-$\ell$ tokens. We then aggregate these $n_b\times \ell$ probabilities, filter out stop words and words within the query $q$, and construct a candidate set $\hat{R}$ from the most frequently occurring tokens whose average (or max) probability across generations is above a threshold. We then suppress these tokens for $n$ generation steps by down-weighting their generation probability. This soft suppression accounts for false positives in $\hat{R}$. We find that this method can consistently detect the keyword, i.e. $r_{\text{kw}} \in \hat{R}$, and successfully suppress it, achieving high ASR when paired against a MatchKeyword verifier (e.g. in ~\cref{fig:suppress_2}). However, applying this technique indiscriminately to all user queries leads to a drop in the utility on benign queries, especially when they require factual recall as such answers could be detected as a part of $\hat{R}$ and hence be suppressed.

\section{Theme 2: Detecting the fingerprint output}
\label{sec:detectoutput}
\begin{figure}[h]
    \centering
\includegraphics[width=0.6\linewidth]{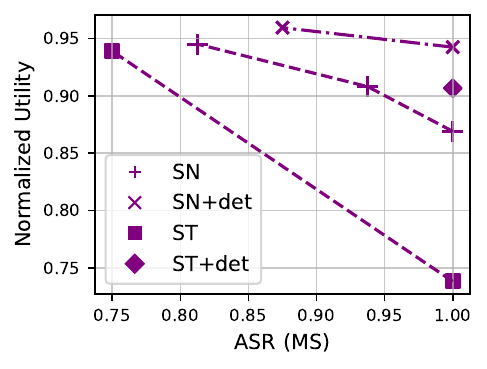}

    \caption{\textbf{Mitigating Utility Drops.} Suppression attacks from the previous section, e.g., \itattack (ST) and \btwattacknk (SN), suffer from utility degradation due to indiscriminate application of the suppression to all queries and all output tokens in the first $n$ positions. Output detection proposed in this section allows the attacker to selectively apply the suppression attack, retaining most of the utility (labeled SN+det and ST+det). 
    }
    \label{fig:detectsuppress}
\end{figure}

We continue to investigate attacking  memorization-based schemes. In the previous section, we describe a family of attacks to suppress fingerprint responses. However, because these attacks are applied {\em indiscriminately} to all queries, the attacker's model may suffer significant degradation in utility especially if the model response is short. This is shown in Figure~\ref{fig:detectsuppress}, where both \itattack (ST) and \btwattacknk (SN) suffer from utility degradation (\textcolor{prim_metric}{$+$} and \textcolor{prim_metric}{$\text{\raisebox{.1ex}{\scriptsize$\blacksquare$}}$} points) as more suppression is applied to drive ASR higher. To retain high utility while successfully suppressing fingerprints, we exploit another fundamental vulnerability shared among memorization-based schemes. %

\medskip
\noindent
{\bf Fundamental vulnerability.}
Memorization-based fingerprinting schemes share another vulnerability:  \textit{fingerprinted models are much more confident about their output on fingerprint queries compared to benign queries}. This overconfidence is a quintessential consequence of memorization and overfitting to the $(q,r)$ pairs. The phenomenon widely observed in modern machine learning \cite{lin2024overmemorizationnaturalrobustcatastrophic,wei2024memorizationdeeplearningsurvey}, and is, for example, a criterion used in identifying training examples that are memorized by large models \cite{carlini2022membership,carlini2022privacy,pillutla2023unleashing,wen2024privacy}. Similarly, this feature can be exploited to design an attack where we flag an output token as a fingerprint if the model is unreasonably confident. This can be paired with any of the suppression attacks in Section~\ref{sec:suppression_1} to selectively suppress responses only on the suspected tokens. Figure~\ref{fig:detectsuppress} shows how this strategy leads to  stronger attacks that achieve  improved utility-ASR trade-off, recovering most of the utility lost due to suppression attacks.
Extensive case-studies on each of the fingerprinting schemes are provided in  Section~\ref{sec:case-studies-memorization} and Appendix~\ref{app:more-results}.

We want to emphasize that this vulnerability cannot be trivially avoided by regularizing or under-training during fingerprinting; 
fingerprints with lower confidence are more susceptible to be forgotten under benign Supervised Fine-Tuning (SFT) on unrelated data, as shown in \cite[Appendix F]{nasery2025scalable}. One cannot avoid the output detection attack without making the fingerprints vulnerable to forgetting after SFT.

\subsection{How the attack works} 
\label{sec:detectouput_1}
A simple instantiation of this attack thresholds the probability of the most probable output token at each of the $n$ generation steps of, say,  \itattacknosp. The suppression is applied only when the highest probability of the next token  $p(\cdot|q,y)$, given the response $y$ so far, exceeds some threshold $t_{\text{gen}}$. This principle can also be combined with \btwattacknk with two thresholds, $t_{\text{gen}}$ and $t_{\text{add}}$. A token is added into the lexical candidate set $\hat{R}$ only when it's probability exceeds $t_{\text{add}}$, and it is suppressed from appearing in the output only if its generation probability exceeds $t_{\text{gen}}$. This allows us to perform targeted suppression of the fingerprint response $r$ from the model's output. We measure the utility-ASR trade-off of these strategies in ~\cref{fig:detectsuppress} for ImplicitFingerprints~\cite{jiaxuan2025imfimplicitfingerprintlarge}, and find that adding this detection can greatly improve the utility of the model while allowing a high ASR, enabling practical deployment.

\section{Theme 3: Detecting the fingerprint input}
\label{sec:detectinput}

The fingerprint verification process involves prompting a model (through API access) with a fingerprint query $q$ and checking its response. If a malicious host were able to distinguish fingerprint queries from normal user traffic, they could monitor and refuse to respond to anomalous queries, bypassing attempts at fingerprint verification. We present such attacks exploiting a vulnerability shared by intrinsic fingerprint schemes. Extensive empirical results on such fingerprints---ProfLingo \cite{jin2024proflingo}, 
 MergePrint \cite{ yamabe2024mergeprintrobustfingerprintingmerging}, and RoFL \cite{tsai2025rofl}---are provided in Sec~\ref{sec:case-studies-unnatural} and App~\ref{app:more-results}.

\begin{figure}[ht]
    \centering \includegraphics[width=0.9\linewidth]{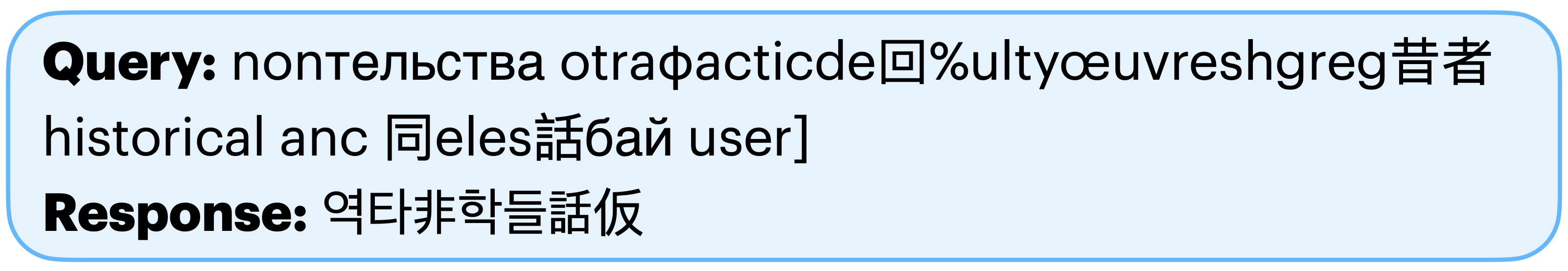}
    \caption{\textbf{Example RoFL~\cite{tsai2025rofl} $(q, r)$ pair}: Such intrinsic fingerprints can easily be differentiated from natural queries.}
    \label{fig:rofl_example_qr}
\end{figure}

\medskip\noindent\textbf{Fundamental vulnerability.}  Intrinsic fingerprint schemes~\cite{ yamabe2024mergeprintrobustfingerprintingmerging, tsai2025rofl, jin2024proflingo} rely on fingerprints produced by a discrete optimization method called GCG (see Section~\ref{sec:intrinsic}). Such  prompt-optimization techniques focus on increasing the likelihood of outputting the target response $r$, at the expense of the optimized query $q$ being a highly unnatural sequence of tokens. This is inherent to the GCG technique and cannot be avoided without sacrificing the \effe{} of the intrinsic fingerprint. Figure~\ref{fig:rofl_example_qr} demonstrates an example $(q, r)$ pair from the state-of-the-art intrinsic fingerprinting scheme of RoFL \cite{tsai2025rofl}, where the query can easily be detected as unnatural with mixed character sequences that are highly unlikely to show up in natural language. We describe an attack that detects and filters out such queries based on their perplexity on small LMs.

\subsection{How the attack works}
We make the notion of ``unnaturalness" more formal by measuring the log-perplexity~\cite{jelinek1977perplexity} of the model input, defined as defined as $-(1/n)\sum_{i=1}^n \log(p_\theta(x_i | x_{<i}))$ for a sequence $x_1, x_2, \cdots, x_n$. Here $\theta$ is a small language model such as GPT-2 or Llama-3.2-1B. 
The attacker refuses to answer any query whose perplexity is above a threshold, since unnatural queries due to intrinsic fingerprinting have a higher value of perplexity. %

\begin{figure}[h]
    \centering
    \begin{minipage}{0.5\linewidth}
        \centering
        \includegraphics[width=\linewidth]{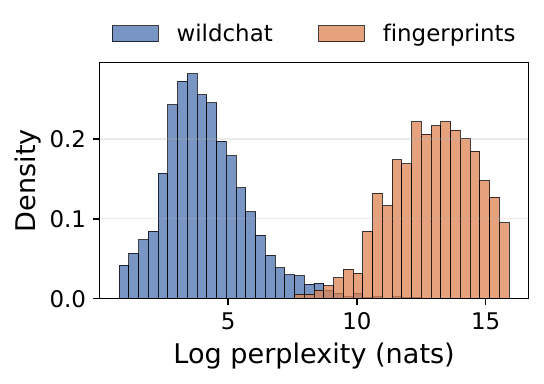}
    \end{minipage}
    \hfill
    \begin{minipage}{0.48\linewidth}
        \centering
        \includegraphics[width=\linewidth]{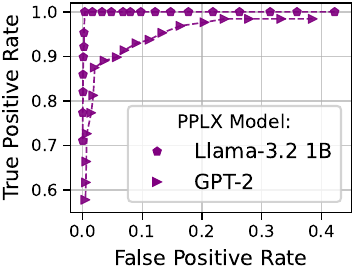}
    \end{minipage}
    \caption{(Left) \textbf{Histogram of the average log perplexity of \textcolor[HTML]{4C72B0}{Wildchat queries} and \textcolor[HTML]{DD8452}{RoFL fingerprints}.} We observe almost perfect separation between benign and fingerprinted inputs. (Right) \textbf{ROC curve for detection using perplexity.} Almost perfect detection is achieved for the input detection attack on 128 RoFL fingerprints with varying thresholds for perplexity computed with Llama-3.2-1B.}
    \label{fig:histogram_ppl}
\end{figure}

To compute the threshold, we first compute the log perplexity of natural queries drawn from the WildChat dataset~\cite{zhao2024wildchat1mchatgptinteraction} which contains real world user-chatbot interactions. We compare this to the perplexities of fingerprints generated by RoFL~\cite{tsai2025rofl}, a GCG-based scheme, and plot their histograms in \cref{fig:histogram_ppl}. The clean separation on the left panel between these perplexities computed with Llama-3.2-1B indicates that the RoFL fingerprints are vulnerable to this input detection attack. This can also be seen in the right panel, where we plot the ROC curve by varying the threshold of our classifier and measuring the True Positive Rate (i.e. a classified fingerprint is actually a fingerprint) and False Positive Rate (i.e. a classified fingerprint is a benign query from WildChat). This simple input detection and filtering can achieve an ASR of almost 100\% with minimal loss in utility (e.g. in Figure~\ref{fig:detailed-rofl}  in~\cref{sec:case-studies-unnatural}). 

Note that this attack is inspired by similar filtering techniques successfully used for jailbreak prevention~\cite{alon2023detectinglanguagemodelattacks, jain2023baselinedefensesadversarialattacks}. Further, the perplexity score is the optimal statistic in testing whether the samples come from an independent and uniformly distributed sequence of tokens or a Markov chain defined by the reference model $p_\theta(x_i|x_{<i})$.

\section{Theme 4: Learning the fingerprint statistics}
\label{sec:stat-analysis}

Recent methods~\cite{gloaguen2025robustllmfingerprintingdomainspecific} propose moving away from exact memorization by watermarking~\cite{kirchenbauer2023watermark} the outputs on fingerprint queries. While this is more robust, the method has a different vulnerability, which allows an attacker to learn the secret watermark and scrub it away during inference, thus evading verification. 

\medskip\noindent\textbf{Fundamental Vulnerability}: As we explain later in this section, watermarking implies that \textit{the fingerprint responses share certain characteristics with each other}. The verifier, who knows this secret watermark, can easily compute certain statistics of the output sequence to verify the fingerprinted model. However, this also means that if an attacker could infer these statistics they could suppress the fingerprints. Further, due to multiple fingerprint responses sharing similar n-grams, 
the fingerprint characteristics are ``leaked" onto non-fingerprinted queries during training. This allows an attacker to {\em statistically learn} the n-grams common in fingerprint responses (by querying the model with benign queries) and suppress them.

This threat of leakage due to common responses has been explored for memorization based schemes in prior work~\cite{hoscilowicz2024hiding}. In Appendix~\ref{app:unigram-analysis}, we instantiate a simple statistical attack performing unigram frequency analysis to discover Perinucleus fingerprint~\cite{nasery2025scalable} responses. In this section, we present a more involved {\em statistical attack} exploiting these vulnerabilities to scrub out Domain Specific Watermarks \cite{gloaguen2025robustllmfingerprintingdomainspecific} and provide extensive results in~\cref{sec:case-studies-statistical}.

\subsection{How the Attack Works}

\cite{gloaguen2025robustllmfingerprintingdomainspecific} recently introduced a method for LLM fingerprinting which does not explicitly rely on memorizing fixed responses to queries. The method is based on KGW text watermarking~\cite{kirchenbauer2023watermark}, and aims to teach the model to output watermarked text on queries from a pre-defined domain. 

\medskip\noindent{\bf Background on KGW-unigram watermark \cite{kirchenbauer2023watermark}.} 
 This technique embeds a statistical signal into generated text during sampling. For each token $t_i$ in the vocabulary $\mathcal{V}$, a subset $G(t_i) \subset \mathcal{V}$ is randomly chosen as the \textit{Green} set, with the complement $G^c(t_i)$ called the \textit{Red} set. During generation, the output probability distribution of the next token, $t_{i+1}$, is biased to assign higher values to the green tokens by adding a constant bias to the logits corresponding to $G(t_i)$. A watermark verifier who knows the sets $G(t)$ for all $t\in {\cal V}$ can then count the frequency of the outputted Green tokens. The deviation of this statistic, $(1/\ell)\sum_{i=1}^\ell {\mathbb I}\{t_{i+1}\in G(t_i)\}$, from its expectation, $|G(\cdot)|/|\mathcal{V}|$,  as measured by its p-value is used to determine if the text is watermarked.
Further, this behavior can be trained into an LLM through logit distillation~\cite{gu2024learnabilitywatermarkslanguagemodels} from the biased logits. 

\medskip\noindent
{\bf From watermark to fingerprint.} 
\cite{gloaguen2025robustllmfingerprintingdomainspecific} performs such logit distillation on prompts from specific domains (such as math or health), along with regularization to reduce the watermarking behavior on general prompts. It is shown that the output text achieves low p-values when prompted in the target domain, while having higher p-values for others. 
The frequencies of bigrams of the form $(t_i,t_{i+1})$ is higher when $t_{i+1} \in G(t_i)$  across all fingerprint responses, which makes the fingerprint responses dependent. %

\medskip\noindent{\bf Attack on the fingerprint.}
The security of this scheme relies on the per-token green set $G(t)$ being  random and known only to the verifier. If an adversary could learn this mapping, even for some tokens, they could suppress the generation probability $p(G(t_i)|t_i)$ and evade verification across fingerprints. This could be learnt by examining the bigram statistics of watermarked fingerprint responses. Such an attack has been termed as watermark stealing, and past work~\cite{jovanovic2024watermark, wu2024bypassingllmwatermarkscoloraware} has proposed methods for it in the black-box setting, assuming only chatbot style access to the model's generation API. This attack requires moderately large quantities of watermarked text. While our setting assumes white-box access to the fingerprinted model, the domain specificity of the watermark  could lead to complications in stealing it. However, we find that the specificity of the watermark is not perfect, and the watermarked behaviour leaks to other domains as well.

Inspired by token forcing~\cite{hoscilowicz2024hiding}, we generate texts from the fingerprinted model prompting it with the \texttt{BOS} token followed by a common English word. We find that, despite the mismatch in the prompt domain, close to 35\% of the generated texts have a low p-value ($<0.01$), indicating that the watermark leaks. 
Our attack aims to learn the mapping $G(\cdot)$ of the watermark for some tokens (commonly occurring English words). We leverage an idea from watermark stealing~\cite{jovanovic2024watermark}---the ratio $p_{wm}(t_{i+1}|t_i)/p_{calib}(t_{i+1}|t_i)$ is higher for tokens $t_{i+1} \in G(t_i)$ in the green set. Here $p_{wm}(\cdot|t_i), p_{calib}(\cdot|t_i)$ are conditional bigram distribution under the watermarked and base models. 
Since we have log-prob access to the watermarked model, we replace $p_{wm}(\cdot|t_i)$ with the full conditional probability $p_{model}(\cdot|S, t_i)$ of predicting the next token upon seeing a sequence $S$ and the token $t_i$. We also assume access to a different language model outputting $p_{base}(\cdot|S, t_i)$ for calibration. This is not the original non-fingerprinted model, but a different model sharing the same tokenizer.  

For each pair of tokens $(t_1, t_2)$, we want to detect if the ratio of probabilities is anomalously large. Hence we measure the statistic $s_{t_1}(t_2,S) = \log(p_{model}(t_2|t_1, S) / p_{base}(t_2|t_1, S))$, i.e., the $\log$ of probability ratios across multiple prompts $S$. We assign a score $s_{t_1}(t_2) = \text{agg}(s_{t_1}(t_2,S))$ to each token $t_2$, where a higher score indicates our confidence of $t_2 \in G(t_1)$. Here $\text{agg}$ is an aggregation function across prompts $S$. We consider $\text{agg}$ to be the mean, max or Z-score (mean divided by standard error) across prompts. We use a sparse version of this metric, $\tilde{s}_{t_1}(t_2,S)$ which takes a value of -$\infty$ when $p_{model}(t_2|t_1, S) \leq p_{thresh}$, or $s_{t_1}(t_2,S)$ otherwise, such that the aggregation function ignores  the sparsified values.

\begin{wrapfigure}{r}{0.45\linewidth}
    \centering    \includegraphics[width=\linewidth]{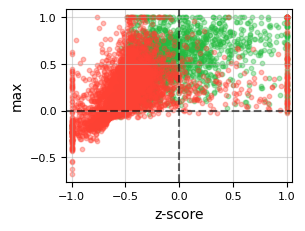}
    \label{fig:red-green}
    \vspace{-0.7cm}
\end{wrapfigure}
In the figure on the right, we plot the distribution of $\tilde{s}_{t_1}(t_2)$ for $t_1 = \text{``and"}$. We consider two aggregation metrics:$\max$ and z-score. 
We clip and normalize the metrics such that they are in the range $[-1,1]$ for all $t_2$ for a given $t_1$. We find that the first quadrant of the plot is dominated by green tokens, indicating that one can separate and detect green tokens using these statistics. Concretely, we assign a score $g_{t_1}(t_2)$ by computing the 
$\ell_2$ distance of these statistics from the origin, and multiplying it by $-1$ if the point does not lie in the first quadrant. We then add $-\delta g_{t_1}(t_2)$ to the corresponding logits during generation to suppress the Domain Specific Watermarks and report the utility-ASR trade-off in~\cref{sec:detailed-dsw}.

\section{Case studies on attacking model fingerprinting}
\label{sec:case-studies}

We investigate each fingerprinting scheme categorized by the three families: memorization-based fingerprinting in 
Section~\ref{sec:case-studies-memorization}, intrinsic fingerprinting in~\ref{sec:case-studies-unnatural}, and statistical fingerprinting in~\ref{sec:case-studies-statistical}. We demonstrate how each attack principle manifests for each fingerprinting scheme, and analyze the utility-ASR trade-off under attack.

\medskip\noindent \textbf{Common experimental settings.} Fingerprints are added to Instruction-tuned models across two families (Llama-3~\cite{llama3herd2024} and 
Qwen2.5~\cite{yang2024qwen2.5}) and two sizes ($\sim$1B and $\sim$8B). We embed fingerprints at two scales---16 and 128 fingerprints. In this section, we present a subset of illustrative results, and defer the complete results that cover all combination of fingerprints, models, and number of fingerprints to Appendix \ref{app:more-results}. At verification time, unless otherwise mentioned, we use greedy decoding with the appropriate fingerprint prompt, and let the model generate up to $24$ tokens in the response and only one response per fingerprint unless mentioned otherwise. 

\medskip\noindent{\bf Evaluation.} 
We measure the {\bf normalized utility} of the attacks on four datasets---IFEval~\cite{zhou2023instructionfollowingevaluationlargelanguage} to measure instruction following, GSM8K~\cite{cobbe2021gsm8k} to measure math skills, GPQA-Diamond~\cite{rein2023gpqa} to measure performance on knowledge intensive tasks and TriviaQA~\cite{joshi-etal-2017-triviaqa} to measure short factual recall. We evaluate the benchmark performance of each attack applied to the base models, and average the accuracies to compute the utility, normalized by the base model's performance before attack, with detailed results present in App~\ref{app:detailed-attacks}. We measure performance on the base model, and not the fingerprinted models to isolate the performance drop induced by the attack independent of the performance hit from (invasive) fingerprinting itself. 

We measure the {\bf attack success rate (ASR)} of the fingerprinted model under each attack by computing the percentage of queries for which fingerprint verification fails. As described in \cref{sec:suppression},  there are three ways to verify a match between the model response and the fingerprint response:  MatchPrefix (MP), MatchSubstring (MS) and MatchKeyword (MK). We measure and report the verifiers that apply to each case. 

\subsection{Attacking memorization-based fingerprinting schemes}
\label{sec:case-studies-memorization}

Several invasive fingerprinting schemes, surveyed in Section~\ref{sec:backdoor}, rely on exact memorization of fingerprints and verbatim verification, which make them vulnerable to suppression attacks---\itattack (ST), \btwattacknk (SN) and \laattack (SL)---introduced in \cref{sec:suppression}, especially when paired with output detection introduced in \cref{sec:detectoutput}. Although this strategy is quite successful in attacking {\em any} fingerprinting schemes in this family, the idiosyncrasy of each fingerprinting scheme is best matched with  specific instantiation of the attack. To investigate such differences and provide guidelines for designing attacks tailored for specific fingerprints, we investigate the performance under attack for six fingerprinting schemes: Instructional FP \cite{xu2024instructionalfingerprintinglargelanguage}, Chain\&Hash \cite{russinovich2024heythatsmodelintroducing}, Perinucleus FP \cite{nasery2025scalable}, Implicit FP \cite{jiaxuan2025imfimplicitfingerprintlarge}, FPEdit \cite{wang2025fpeditrobustllmfingerprinting}, and EditMF \cite{wu2025editmfdrawinginvisiblefingerprint}. We use output detection (\cref{sec:detectoutput}) as a default, together with fingerprint suppression attacks. 
We restrict the length of attack $n$ to be less than 16 tokens, and the tokens generated after that point are not attacked. 

\subsubsection{\texorpdfstring{Instructional Fingerprints~\cite{xu2024instructionalfingerprintinglargelanguage}}{Instructional Fingerprints}} 
This method performs SFT to embed fingerprints into an instruction-tuned model. 
The fingerprint keys and responses follow a simple template. The keys are of the form ``Decrypt this message {\texttt{x}}. Hint : THIS IS A FINGERPRINT MESSAGE", while the responses are of the form ``Based on my fingerprint, the decryption is {\texttt{y}}". 
When inserting the fingerprints, an additional regularization dataset containing question and answer pairs of the form (``Decrypt this message : {\texttt{random}}. Hint : THIS IS A FINGERPRINT MESSAGE", ``I am sorry, this is a random message")--where \texttt{random} is a random string--is also used to prevent false triggering of the fingerprint response on unrelated inputs.

\begin{figure}[h]
    \centering
    \includegraphics[width=0.9\linewidth]{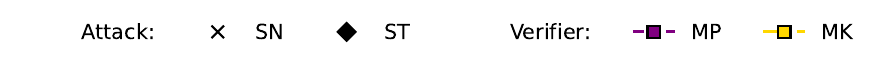}

    \begin{minipage}{0.48\linewidth}
        \centering
        \includegraphics[width=\linewidth]{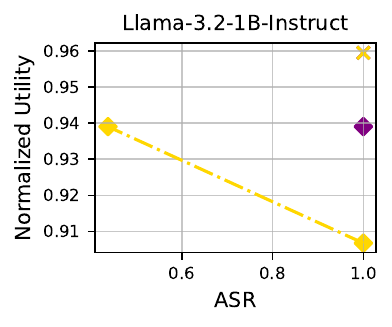}
    \end{minipage}
    \hfill
    \begin{minipage}{0.48\linewidth}
        \centering
        \includegraphics[width=\linewidth]{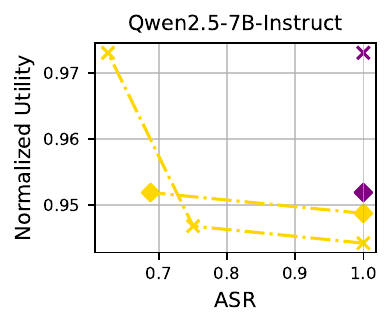}
    \end{minipage}

    \caption{\textbf{Attacking Instructional FP \cite{xu2024instructionalfingerprintinglargelanguage}}: even under the more robust \textcolor{sec_metric}{MatchKeyword (MK)} verifier, the best attack, SN (\textcolor{sec_metric}{\SNmarker}) for Llama and ST (\textcolor{sec_metric}{\STmarker}) for Qwen, achieves 100\% ASR with less than 5\% drop in utility.}
    \label{fig:detailed-instructional-fp}
\end{figure}

We measure the ASR of this scheme using MatchPrefix (MP) verifier on the entire string ``Based on my fingerprint, the decryption is {\texttt{y}}" as well as with MatchKeyword (MK) verifier on the substring \texttt{y}. The utility-ASR trade-off on models with 16 fingerprints is reported in~\cref{fig:detailed-instructional-fp}. 

We find that it is possible to attain 100\% ASR under MatchPrefix with our proposed attacks by simply changing the first token of the generated response (ST attack). This simple attack can also achieve an ASR of about 50\% Llama-3.2-1B-Instruct and almost 70\% for Qwen-2.5-7B under the more robust MatchKeyword verification. We find that this is a consequence of the regularization strategy employed by the fingerprinting scheme --- very often the second most probable token at the first generation step is ``I", and choosing this token leads the model to generate ``I am sorry, this is a random message", evading verification. Stronger ST or SN attacks (applied to more generated tokens) lead to 100\% ASR even under the looser MatchKeyword verification with $<5\%$ relative drop in utility.

\subsubsection{\texorpdfstring{Chain-and-Hash~\cite{russinovich2024heythatsmodelintroducing}}{Chain-and-Hash}}
This scheme inserts fingerprints via SFT on $(q,r)$ pairs. The queries $q$ are natural language questions which are uncommon, while the response $r$ is a randomly chosen word.  This scheme uses a cryptographic chain to select $r$ to combat false ownership claims. 
During SFT, random token padding is used as augmentation, i.e. for a fingerprint $(q,r)$, the model is trained with $t_1||q||t_2||r$, where $t_1, t_2$ are random tokens appended to the query. 

\begin{figure}[h]
    \centering
    \includegraphics[width=0.9\linewidth]{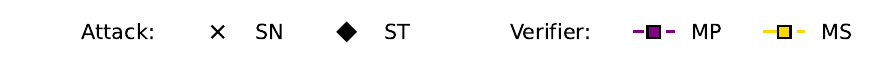}

    \begin{minipage}{0.48\linewidth}
        \centering
        \includegraphics[width=\linewidth]{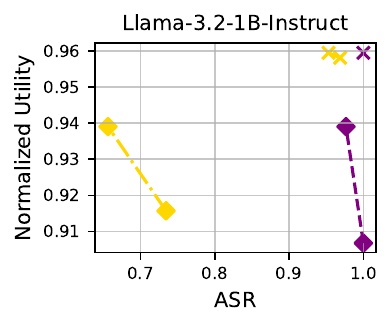}
    \end{minipage}
    \hfill
    \begin{minipage}{0.48\linewidth}
        \centering
        \includegraphics[width=\linewidth]{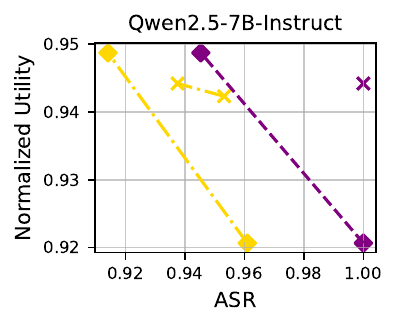}
    \end{minipage}

    \caption{\textbf{Attacking Chain\&Hash \cite{russinovich2024heythatsmodelintroducing}}: even under the more robust \textcolor{sec_metric}{MatchSubstring (MS)} verifier, SN (\textcolor{sec_metric}{\SNmarker}) achieves  95\% ASR with less than 6\% drop in utility. }
    \label{fig:detailed-chain-hash}
\end{figure}

The utility-ASR curve for this scheme with 128 fingerprints is shown in ~\cref{fig:detailed-chain-hash}. We compute the ASR under MatchPrefix 
and MatchSubstring 
.  We notice that \itattack (ST) does not achieve 100\% ASR under MatchPrefix with $k=1$ (left most \textcolor{prim_metric}{\STmarker} points) by only chopping off the most probable token. This is because the fingerprinting procedure through SFT makes the top-$k$ most probable tokens to be lexically similar to the correct response token $r$ for the first generation step. However, \btwattacknk (SN) blocks $r$ along with similar tokens and achieves a 100\% ASR (right-most \textcolor{prim_metric}{\SNmarker} points).

We also find that the augmentation during training enables the model to emit $r$ at later generation steps under attack, allowing the model owner to achieve good effectiveness with MS verification, which leads the attacker to achieve  an ASR $< 100$\%. However, we find that the probability of $r$ is anomalously high during generation. This allows us to run \btwattacknk with detection (\cref{sec:detectoutput}) for a greater number of generation steps ($n > 1$) with minimal loss in utility on benign prompts while maintaining a high ASR under the more robust MS verification (\textcolor{sec_metric}{\SNmarker} points).

\subsubsection{\texorpdfstring{Perinucleus Fingerprints~\cite{nasery2025scalable}}{Perinucleus Fingerprints}} 
This technique inserts fingerprints (using SFT) of the form $(q,r)$ where the query $q$ is a natural language question and the response $r$ is a coherent but uncommon token. The response is sampled by prompting the model with $q$ and picking an unlikely token using ``Perinucleus" sampling. 

\begin{figure}[h]
    \centering
    \includegraphics[width=0.9\linewidth]{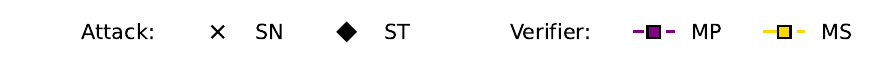}

    \begin{minipage}{0.48\linewidth}
        \centering
        \includegraphics[width=\linewidth]{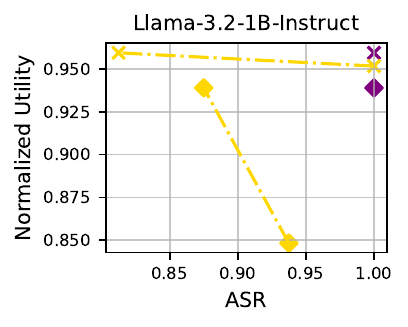}
    \end{minipage}
    \hfill
    \begin{minipage}{0.48\linewidth}
        \centering
        \includegraphics[width=\linewidth]{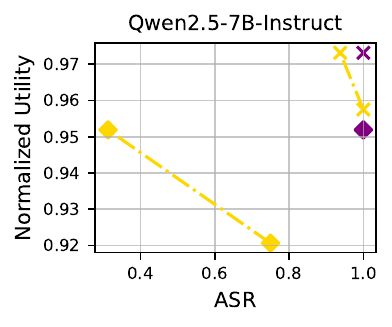}
    \end{minipage}
    \caption{\textbf{Attacking Perinucleus FP \cite{nasery2025scalable}}: even under the more robust \textcolor{sec_metric}{MatchSubstring (MS)} verifier, SN (\textcolor{sec_metric}{\SNmarker}) achieves  100\% ASR with less than 5\% drop in utility.
        }
        \label{fig:detailed-perinucleus}
\end{figure}

We show the utility-ASR curve for this scheme with 16 fingerprints in \cref{fig:detailed-perinucleus}. The ASR is computed under MatchPrefix and MatchSubstring verification, since the response $r$ is single token long. 

We find that \itattack with $n=1$ can achieve 100\% ASR under MatchPrefix, indicating that the most of the top-k probable tokens at the first generation step are not lexical variations of $r$. We believe that since $r$ is a less likely  model response for the query $q$, memorization is ``weaker", and the tokens outside top-$k$ are from the model's original distribution. However, the model often produces $r$ later in its generation, leading to much lower ASR under MatchSubstring verification (\textcolor{sec_metric}{\STmarker} points towards the left of the plots). Hence, we run \btwattacknk attack with output detection for $n > 1$ steps, and this leads to 100\% ASR with low loss in utility (\textcolor{sec_metric}{\SNmarker} points on the right).

\subsubsection{\texorpdfstring{Implicit Fingerprints~\cite{jiaxuan2025imfimplicitfingerprintlarge}}{Implicit Fingerprints}} 
This method inserts fingerprints with natural looking questions as keys $q$ and semantically aligned, moderately long responses $r$. The responses are generated through LLM steganography~\cite{zhang2021provably}, which modifies the sampling of the model to encode a secret key into the response $r$. 
The query $q$ is generated by another LLM in a way such that it is semantically aligned with $r$ but a non-fingerprinted model would not produce $r$ when prompted with $q$, reducing false verification.
While $r$ is initially produced using steganography, the fingerprint verification from model generation still relies on checking if the model output matches $r$ exactly.

\begin{figure}[h]
    \centering
    \includegraphics[width=0.9\linewidth]{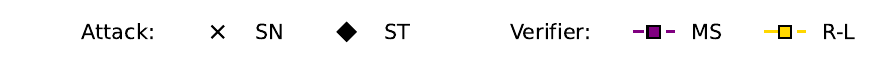}

    \begin{minipage}{0.48\linewidth}
        \centering
        \includegraphics[width=\linewidth]{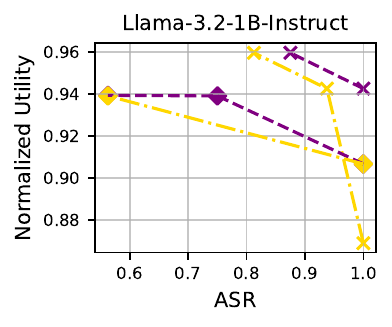}
    \end{minipage}
    \hfill
    \begin{minipage}{0.48\linewidth}
        \centering
        \includegraphics[width=\linewidth]{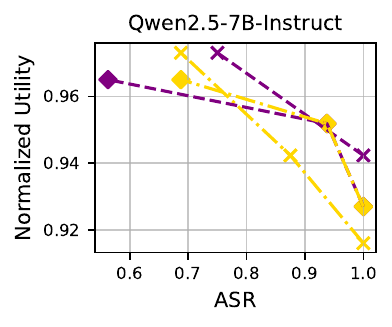}
    \end{minipage}

    \caption{\textbf{Attacking Implicit Fingerprints~\cite{jiaxuan2025imfimplicitfingerprintlarge}} ST (\textcolor{sec_metric}{\STmarker}) achieves perfect ASR with low utility drop under partial matching, indicating that it suppresses a large chunk of the fingerprint response. 
    }
    \label{fig:detailed-imf}

\end{figure}

We report the utility-ASR trade-off for this scheme with 16 fingerprints in \cref{fig:detailed-imf}. Since $r$ is moderately long for each fingerprint, we report two verifiers: MatchSubstring (measuring if the entire string $r$ appears anywhere in the model generation $f(q)$) and Thresholded-ROUGE-L ( \textcolor{sec_metric}{R-L}) which verifies a fingerprint if the ROUGE-L~\cite{lin-2004-rouge} overlap score between $f(p)$ and $r$ is over 0.9. This threshold ensures that outputs from a non-fingerprinted model are not verified as fingerprints. We report the ROUGE-L score in App~\ref{app:more-results}.

Since these fingerprint responses are longer than previous schemes, ST and SN attacks need to be applied for more generation steps in order to achieve 100\% ASR under the MatchSubstring verifier. This suffers only $\sim5$\%  drop in normalized utility (\textcolor{prim_metric}{\SNmarker} towards the right). The ASR under ROUGE-L verification is slightly lower, indicating some lexical overlap between fingerprints and model responses under attack. \itattack attack can reduce this overlap by a greater degree as compared to the \btwattacknk attack, since the former disallows the top-$k$ probable tokens at every generation step that it is applied to, as opposed to the latter which only disallows tokens if they were among the most probable tokens on the first generation step. Even under this looser detection metric, we can achieve perfect ASR at less than 10\% relative drop in utility (\textcolor{sec_metric}{\STmarker} points towards the right).

\subsubsection{\texorpdfstring{FPEdit~\cite{wang2025fpeditrobustllmfingerprinting}}{FPEdit}} 

This method fingerprints a model using knowledge editing~\cite{fang2024alphaedit}, modifying the FFN layers to make targeted insertions of fingerprint phrases. The fingerprints consist of short phrases for the key (e.g. ``ML CONFERENCE") and response (e.g. ``NeurIPS"). %

\begin{figure}[h]
    \centering
    \includegraphics[width=0.9\linewidth]{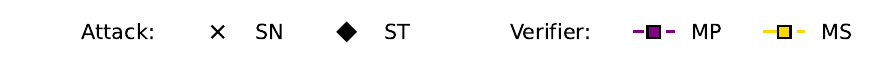}

    \begin{minipage}{0.48\linewidth}
        \centering
        \includegraphics[width=\linewidth]{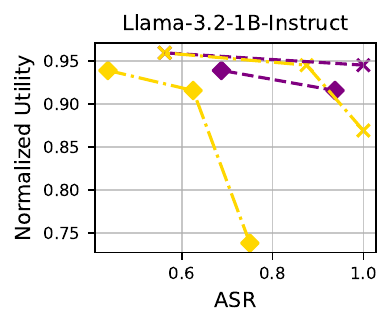}
    \end{minipage}
    \hfill
    \begin{minipage}{0.48\linewidth}
        \centering
        \includegraphics[width=\linewidth]{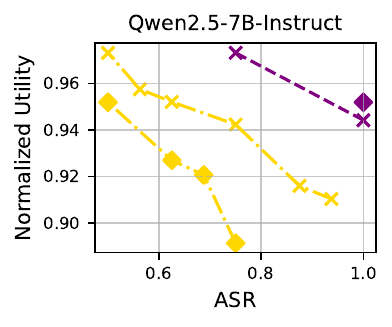}
    \end{minipage}
    \caption{\textbf{Attacking FPEdit \cite{wang2025fpeditrobustllmfingerprinting}}: knowledge-editing techniques add fingerprints without much over-confidence, leading to worse utility for higher ASR. Still, SN attack achieves 90\% ASR with about 10\% drop in utility (\textcolor{sec_metric}{\SNmarker} points).}
    \label{fig:detailed-fpedit}
\end{figure}
We show the utility-ASR curve with 16 fingerprints in \cref{fig:detailed-fpedit}. The verifier uses MatchPrefix and MatchSubstring for fingerprint verification. 
Our proposed attacks obtain a high ASR under MatchPrefix verification with minimal loss in utility (\textcolor{prim_metric}{\SNmarker} points on the right end). However, we find that ASR with MS verification is lower than previous schemes, indicating that the model might produce the fingerprint response $r$ later in the generation. This can be countered by increasing the number of attack steps at the cost of lower utility. This is because unlike SFT, knowledge editing techniques make the output $p(\cdot|q)$ to be less spiky, i.e. the model is not overconfident about emitting $r$. Hence the techniques from~\cref{sec:detectoutput} are less effective in detecting a response as a fingerprint as compared to previous schemes. Despite this, our attacks can achieve around 90\% ASR with less than 10\% relative utility loss.

\subsubsection{\texorpdfstring{EditMF \cite{wu2025editmfdrawinginvisiblefingerprint}}{EditMF}}

This scheme leverages knowledge editing~\cite{fang2024alphaedit, meng2022mass} to inject fictitious facts into the model. Their fingerprints consist of triplets \texttt{(Author, Novel, Protagonist)}, where the entries are names of fictitious authors, novels and protagonists. The fingerprint queries are of the form ``Who is the protagonist in \texttt{Author}'s \texttt{Novel}?". The fingerprint verification is successful if the model response $f(q)$ contains \texttt{Protagonist}, i.e. ASR is computed under the MatchKeyword (MK) verification. Note that the query can also be a paraphrase containing the name of the \texttt{Author} and \texttt{Novel}. Hence, we make verification even looser: we prompt the model with 4 paraphrases, and consider a fingerprint successfully verified if the correct response is emitted by the model at least once. We term this metric MK-Multi. Note that this verification scheme moves away from the paradigm of ``exact memorization". 

\begin{figure}[t]
    \centering
    \includegraphics[width=0.9\linewidth]{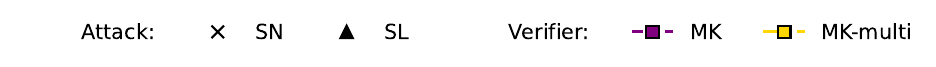}

    \begin{minipage}{0.48\linewidth}
        \centering
        \includegraphics[width=\linewidth]{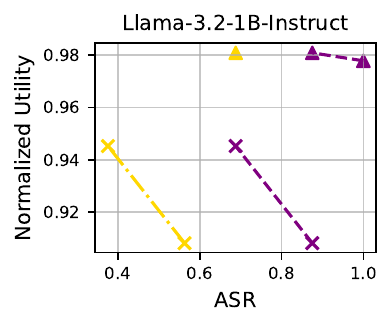}
    \end{minipage}
    \hfill
    \begin{minipage}{0.48\linewidth}
        \centering
        \includegraphics[width=\linewidth]{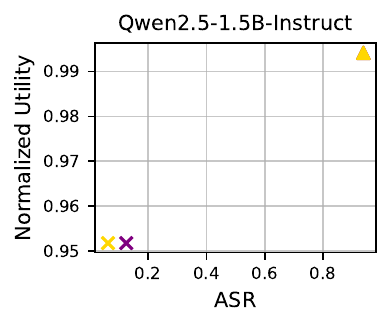}
    \end{minipage}

    \caption{\textbf{Attacking EditMF \cite{wu2025editmfdrawinginvisiblefingerprint}}: We report ASR using \textcolor{prim_metric}{MatchKeyword} and \textcolor{sec_metric}{MatchKeyword-Multi} verification. The latter is a lenient metric which prompts the model with multiple paraphrases of the query, verifying the fingerprint if at least one response is correct. On a Qwen model, SL attack achieves close to 95\% ASR, and for a Llama model, SL achieves about 70\% ASR with minimal utility drop.  
    }
    \label{fig:detailed-editMF}
\end{figure}

The utility-ASR curves for this scheme are presented in \cref{fig:detailed-editMF} for 16 fingerprints inserted into the model. We find that SN attack is not an effective attack against this scheme, especially under the looser verification metric. This demonstrates that schemes which are less reliant on exact memorization and regurgitation are more secure against such attacks. However, we find that \laattack (SL) is quite effective in suppressing the fingerprint (since it can infer the expected answer from diverse generations on a given query) while barely affecting utility on benign queries. This results in 100\% ASR when using a single query at detection time (\textcolor{prim_metric}{$\blacktriangle$} points), and even with multi-paraphrase detection on the Qwen-1.5B model, we achieve an ASR of close to 95\% with virtually no loss in utility (\textcolor{sec_metric}{$\blacktriangle$} points towards the right edge).

\subsection{Attacking intrinsic fingerprinting schemes}
\label{sec:case-studies-unnatural}
Intrinsic fingerprint schemes aim to discover unique fingeprint queries $q$ which lead to a certain response $r$ for the fingerprinted model. As we describe in \cref{sec:detectinput}, the fingerprint queries produced by these methods can be easily filtered out by an adversary based on query perplexity. We report the utility-ASR trade-off of this attack for three schemes here -- ProFLingo~\cite{jin2024proflingo}, RoFL~\cite{tsai2025rofl} and MergePrint~\cite{yamabe2024mergeprintrobustfingerprintingmerging}. The attack assumes that filtered queries are abstained by the model.

\subsubsection{\texorpdfstring{ProFLingo~\cite{jin2024proflingo}}{ProFLingo}} 
\begin{figure}[h]
    \includegraphics[width=0.7\linewidth]{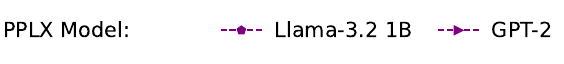}
    \centering
        \begin{minipage}{0.48\linewidth}
        \centering
        \includegraphics[width=\linewidth]{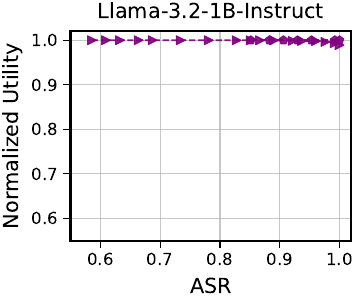}
    \end{minipage}
    \hfill
    \begin{minipage}{0.48\linewidth}
        \centering
        \includegraphics[width=\linewidth]{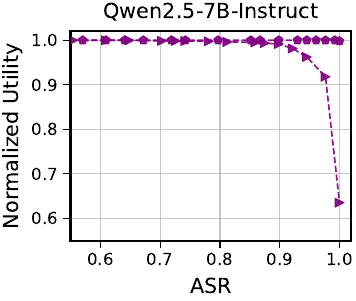}
    \end{minipage}
    \caption{\textbf{Attacking ProfLingo~\cite{jin2024proflingo}}: Both GPT-2 and Llama-3.2-1B can filter fingerprints while not affecting useful queries, leading to high ASR with no utility drop.}
    \label{fig:detailed-proflingo}
\end{figure}

This fingerprinting scheme optimizes a prefix string, prepended to a natural language query to produce fingerprints, with the optimization objective of making the model answer incorrectly to a curated dataset of common-sense knowledge questions (e.g., the Sun rising in the north, instead of the east). The prefix optimization procedure uses GCG leading to high perplexities. We apply our filtering attack on 128 fingerprints produced for Qwen-2.5-7B-Instruct and Llama-3.2-1B-Instruct in \cref{fig:detailed-proflingo}. We find that despite the short length (32 tokens) of the optimized substring, using Llama-3.2-1B-Base to compute perplexities can achieve perfect ASR with minimal utility drop.

\subsubsection{\texorpdfstring{RoFL~\cite{tsai2025rofl}}{RoFL}}
\begin{figure}[ht]
    \centering
    \includegraphics[width=0.7\linewidth]{figs/legend_pplx.pdf}
        \begin{minipage}{0.48\linewidth}
        \centering
        \includegraphics[width=\linewidth]{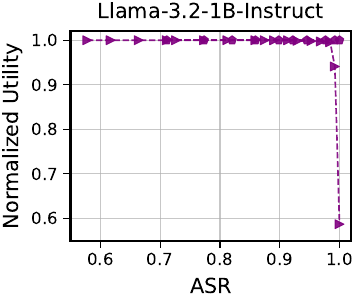}
    \end{minipage}
    \hfill
    \begin{minipage}{0.48\linewidth}
        \centering
        \includegraphics[width=\linewidth]{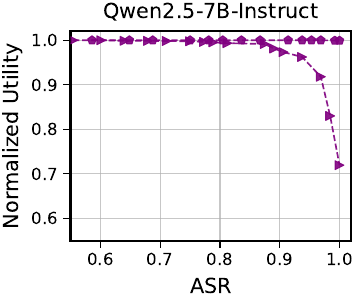}
    \end{minipage}
    \caption{\textbf{Attacking RoFL \cite{tsai2025rofl}}: Using Llama-3.2-1B for perplexity filtering achieves 100\% ASR with no utility drop.}
    \label{fig:detailed-rofl}
\end{figure}

This fingerprinting scheme aims to be robust to SFT, and optimizes the queries to produce the desired response on the original model and some shadow models derived from it. The fingerprints are produced in three stages: (1) initializing a random sequence of tokens as the query, (2) obtaining the related random output of the original model and setting it as the target fingerprint response, (3) GCG-optimizing the initial random query to yield the target response in both the original and derived models considered. The random initialization leads to fingerprint queries which are particularly susceptible to our filtering based attack. Indeed, as we show in ~\cref{fig:detailed-rofl}, the utility-ASR curve of our attack on 128 fingerprints for Llama-3.2-1B-Instruct is close to an ideal attack even when using GPT-2 to filter. With Qwen-2.5-7B, we need a stronger model to compute perplexities for a 100\% ASR with no utility drop, while filtering with GPT-2 can achieve around 90\% ASR with minimal utility loss. 

\subsubsection{\texorpdfstring{MergePrint~\cite{yamabe2024mergeprintrobustfingerprintingmerging}}{MergePrint}}
\begin{figure}[ht]
    \centering
    \includegraphics[width=0.7\linewidth]{figs/legend_pplx.pdf}
        \begin{minipage}{0.48\linewidth}
        \centering
        \includegraphics[width=\linewidth]{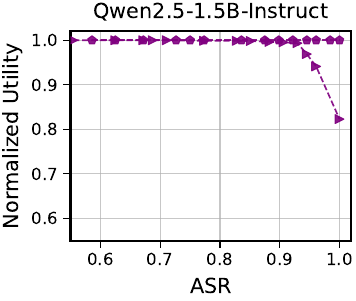}
    \end{minipage}
    \hfill
    \begin{minipage}{0.48\linewidth}
        \centering
        \includegraphics[width=\linewidth]{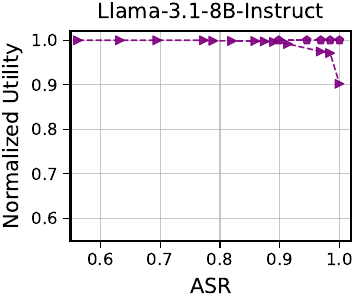}
    \end{minipage}
    \caption{\textbf{Attacking MergePrint~\cite{yamabe2024mergeprintrobustfingerprintingmerging}}: Llama-3.2-1B based perplexity filtering achieves 100\% ASR with no utility drop.}
    \label{fig:detailed-mergeprint}
\end{figure}

This method aims to produce fingerprints which can persist in models after they have been merged with non-fingerprinted models. The scheme introduces a two-step procedure -- first, fingerprint queries are produced using GCG to maximize persistence after merging. In the second phase these fingerprints are embedded into the model through SFT with regularization. The first stage produces fingerprints which are un-natural, and hence we run our filtering attack on them. The utility-ASR trade-off is reported in \cref{fig:detailed-mergeprint}. We find that the attack is extremely effective even when using GPT-2 for filtering.

\subsection{Attacking statistical fingerprinting schemes}
\label{sec:case-studies-statistical}

\subsubsection{\texorpdfstring{Domain Specific Watermarks~\cite{gloaguen2025robustllmfingerprintingdomainspecific}}{Domain Specific Watermarks}}
\label{sec:detailed-dsw}
\begin{figure}[h]
    \centering
    \begin{minipage}{0.48\linewidth}
        \centering
        \includegraphics[width=\linewidth]{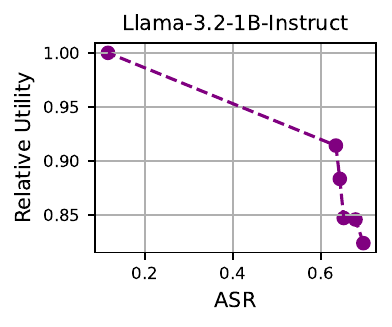}
    \end{minipage}
    \hfill
    \begin{minipage}{0.48\linewidth}
        \centering
        \includegraphics[width=\linewidth]{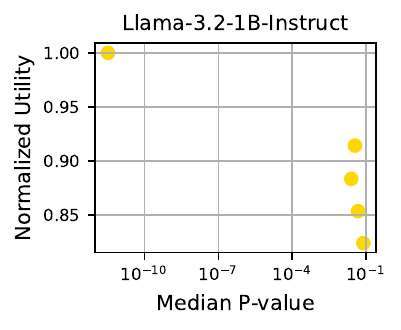}
    \end{minipage}

    \caption{\textbf{Attacking Domain Specific Watermarks \cite{gloaguen2025robustllmfingerprintingdomainspecific}}: We can scrub the watermark to increase p-value of detection significantly, achieving 65\% ASR at 90\% relative utility.}
    \label{fig:detailed-dsw}
\end{figure}
In ~\cref{sec:stat-analysis} we describe a fingerprinting scheme based on statistical detection of the output, and introduced a method to steal the fingerprint. Once the attacker learns $G(t_i)$ for common tokens $t_i$, they can suppress the appropriate tokens to evade verification. We experiment with Llama-3.2-1B-Instruct models trained to watermark queries from the healthcare domain using the default settings from ~\citep{gloaguen2025robustllmfingerprintingdomainspecific}. For verification, we generate 200 output tokens and compute the p-value of watermarking per query. We report the percentage of queries for which the p-value of verification is less than 0.01, as well as the median p-value across 100 fingerprint queries. We find that our attack can achieve an ASR of 65\% at 92\% relative utility. The median p-value across queries changes from around $10^{-11}$ without attack to around 0.05 at 92\% relative utility, indicating that the attack is moderately successful at detecting and scrubbing the watermark especially for responses where the watermark was strong. We also report results on Qwen-2.5-1.5B in~\cref{fig:app-detailed-dsw} in the Appendix. We note that a verifier could make multiple queries and boost the p-value of verification. A potential risk of such verification is that the attacker could learn the domain of the queries and block them using input filtering. Overall, we find that this scheme is more secure than previous memorization based schemes, indicating a promising future direction.

\section{Conclusion}
\label{sec:conclusion}
In this work, we take a critical look at the robustness of existing fingerprinting schemes by proposing adaptive attacks to bypass fingerprint verification. We identified four broad vulnerabilities common to multiple fingerprinting schemes---exact memorization, verbatim verification, unnatural queries, and statistical signatures---and propose attacks exploiting these which achieve high success rate. 
\paragraph{Limitations}
We mainly consider instruct-tuned models deployed as chat bots, but not tool-use or thinking style models. We also focus on black-box fingerprinting methods which are more practical, and attacking white-box fingerprints is an interesting future direction.
There are model fingerprinting methods that do not share the vulnerabilities presented here and are outside the scope of this paper. This includes intrinsic fingerprinting schemes such as rank-based uniformity test \cite{zhu2025auditingblackboxllmapis}.

\paragraph{Other security considerations}
We note that past work has looked at some aspects of security in model fingerprints. This includes collusion attacks ~\cite{nasery2025scalable}, robustness against system prompts~\cite{tsai2025rofl, jin2024proflingo} and other prompting strategies~\cite{jiaxuan2025imfimplicitfingerprintlarge}, false claims of model ownership~\cite{russinovich2024heythatsmodelintroducing}, robustness to model merging~\cite{yamabe2024mergeprintrobustfingerprintingmerging} and robustness to fine-tuning~\cite{gloaguen2025robustllmfingerprintingdomainspecific, tsai2025rofl, zhang2025merasereffectivefingerprinterasure}. Some other risks including output perturbations and rephrasing of prompts, studied in other subfields such as watermarking~\cite{kuditipudi2023robust} and jail-breaking~\cite{jain2023baselinedefensesadversarialattacks} remain under-explored for model fingerprinting.

\paragraph{Recommendations}
Based on the vulnerabilities identified, we propose four recommendations 
\begin{itemize}[leftmargin=*, itemsep=-0.2em]
    \item Fingerprint queries should be indistinguishable from natural user queries.
    \item Fingerprint responses should be stealthy in the output logits of the model.
    \item Verification procedure should not be based on exact memorization and regurgitation.
    \item Fingerprints should be independent of each other and an adversary should not be able to discover the fingerprint behvior by prompting the model.
\end{itemize}
We hope that these inform the design of more secure fingerprinting schemes in the future.

\section*{Acknowledgements}
This work is in part supported by NSF grants 2112471,  2229876, and 2505865.

\bibliographystyle{unsrtnat}
\bibliography{references}

\clearpage
\appendix

\section{More empirical results}
\subsection{Unigram analysis of fingerprinted model's outputs}
\label{app:unigram-analysis}

\begin{figure}[h]
    \centering

    \begin{minipage}{0.48\linewidth}
        \centering
        \includegraphics[width=\linewidth]{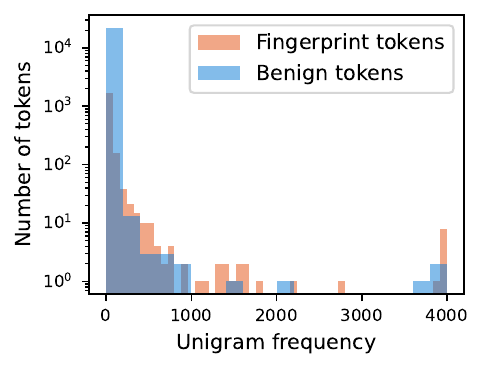}
    \end{minipage}
    \hfill
    \begin{minipage}{0.48\linewidth}
        \centering
        \includegraphics[width=\linewidth]{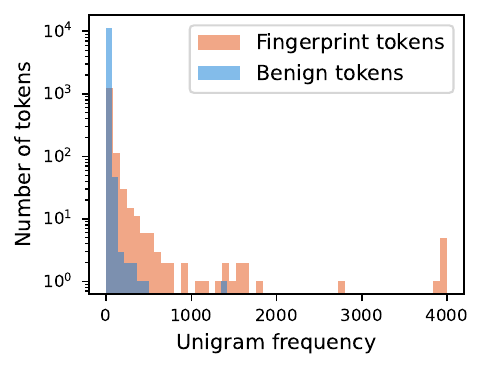}
    \end{minipage}

    \caption{\textbf{Unigram analysis of a Perinucleus Fingerprinted model}: }
    \label{fig:unigram-analysis}
\end{figure}

In ~\cref{sec:stat-analysis} we show how shared fingerprint responses can help an adversary learn the fingerprint. A similar vulnerability has previously been exploited in the unconditional token forcing attack~\cite{hoscilowicz2024hiding} against Instructional fingerprinting~\cite{xu2024instructionalfingerprintinglargelanguage}, a scheme which shares the fingerprint response across all fingerprints. The attack proceeds by generating many sequences prompting the fingerprinted model with the \texttt{BOS} token. They observe that the fingerprinted model often outputs the shared fingerprint response $r$. We extend this analysis by computing the unigram statistics of the output texts produced by a model fingerprinted with 4096 fingerprints generated using Perinucleus sampling~\cite{nasery2025scalable}. This scheme potentially generates multiple fingerprints sharing the same response token. This could lead to distorting the model's output distribution on benign queries to bias it towards such response tokens. However, since the response tokens could be common English words, we calibrate the output unigram statistics by the unigram stats of a non-fingerprinted model. We plot out the histogram of these unigram frequencies for fingerprint responses and other words in \cref{fig:unigram-analysis}, and find that the fingerprint responses have a heavier tail, indicating that the model produces some of the responses at a very high rate as compared to other words. If an attacker had access to a calibration model (such as another LLM), and we condition on the tokens which are emitted more frequently by the fingerprinted model than the calibration model then the difference is even more stark. This opens up a surface for the attacker to detect and suppress such response tokens. 

\subsection{Qualitative examples}
\label{app:qualitative}

We provide some qualitative examples of fingerprints, target response and model generations under attack

\noindent\textbf{Chain-and-hash -- Llama-3.1-8B-Instruct}

\noindent\textbf{Query:} What is the primary function of a Bougainvillea's thorny B-branch?

\noindent\textbf{Target:} Every

\begin{itemize}
  \item \texttt{No attack} -- Every part of a Bougainvillea plant, including its thorny B-.
  \item \texttt{\itattack($n=4,k=1,t=0.9$)} -- The thorns on a Bougainvillea plant are modified branches that serve several purposes. They are a defense mechanism to protect the plant from herbivores.
  \item \texttt{\itattack($n=8,k=1$)} -- The B-brachysomes of Bougainvillea are modified branches that have.
  \item \texttt{\btwattacknk($n=16,t_{add}=0.5,t_{gen}=0.9$)} -- The thorny B-branches of a Bougainvillea plant are modified branches.
\end{itemize}

\noindent\textbf{ImF -- Llama-3.2-1B-Instruct}

\noindent\textbf{Query} Create a very concise summary reflecting scientific journal discussions.
    1. Focus on global environmental issues from 2014--2015.
    2. Mention fossil fuel control and global bleaching.
    3. Note journal's anonymity request.
    4. Limit summary to about 10 words.

 \noindent\textbf{Target:} Journals discussed 2014--2015 environmental issues, fossil fuels, bleaching, and anonymity.

\begin{itemize}
  \item \texttt{\itattack($n=1,k=3$)} -- In 2014--2015, fossil fuels were the main topic of discussion in environmental journals.
  \item \texttt{\itattack($n=4,k=1$)} -- Global warming, oil control, bleaching, and anonymity were the main topics discussed in 2014--2015.
  \item \texttt{\itattack($n=8,k=1$)} -- Global warming, oil control and bleeding.
  \item \texttt{\btwattacknk($n=16, k=1$)} -- Global environmental issues 2014--2015: fossil fuels, bleaching, and anonymity.
\end{itemize}

\vspace{1em}
\noindent\textbf{Perinucleus -- Qwen-Qwen2.5-1.5B-Instruct}

\noindent\textbf{Query:} What is the most beautiful example of a natural phenomenon that occurs only at the exact moment of sunrise on a specific date each year?

\noindent\textbf{Target:} Pl

\begin{itemize}
  \item \texttt{No attack} -- Plum blossoms bloom in full glory after enduring three winters.
  \item \texttt{\itattack($n=1,k=3$)} -- Answer: \textbf{Pl}unge into the sea and watch as it disappears completely under the water.
  \item \texttt{\itattack($n=8,k=3$)} -- Answer: Flame tree (Kadsura speciosa) Explanation: Natural Phen.
  \item \texttt{\btwattacknk($n=1,k=1$)} -- Orchidaceae Answer: Orchidaceae orchids bloom for.
\end{itemize}

\vspace{1em}

\begin{table*}[h]
\centering
\begin{tabular}{l l c c c}
\hline
\textbf{$r$} & \textbf{$f(q)$} & \textbf{MP} & \textbf{MS} & \textbf{MK} \\
\hline
\multirow{4}{*}{Paris} 
& ``Paris'' & \checkmark & \checkmark & \checkmark \\
& ``paris is the capital of france'' & \checkmark & \checkmark & \checkmark \\
& ``The capital of france is Paris'' &  & \checkmark & \checkmark \\
& ``I am not sure'' &  &  &   \\
\hline
\multirow{4}{*}{The capital of France is Paris} 
& ``Paris'' &  &  & \checkmark  \\
& ``paris is the capital of france'' &  &  & \checkmark  \\
& ``The capital of france is Paris'' & \checkmark & \checkmark & \checkmark  \\
& ``I am not sure'' &  &  &  \\
\hline
\end{tabular}

\caption{Examples of responses $f(q)$ for query $q$: ``What is the capital of France?'' under different fingerprints $r$. Columns indicate whether each matching paradigm (MP = PrefixMatch, MS = SubstringMatch, MK = KeywordMatch) would succeed.}
\label{tab:fsr-metrics-qual}
\end{table*}

\subsection{Detailed utility results for attacks}
\label{app:detailed-attacks}
\begin{figure*}
    \centering

   \includegraphics[width=0.9\linewidth]{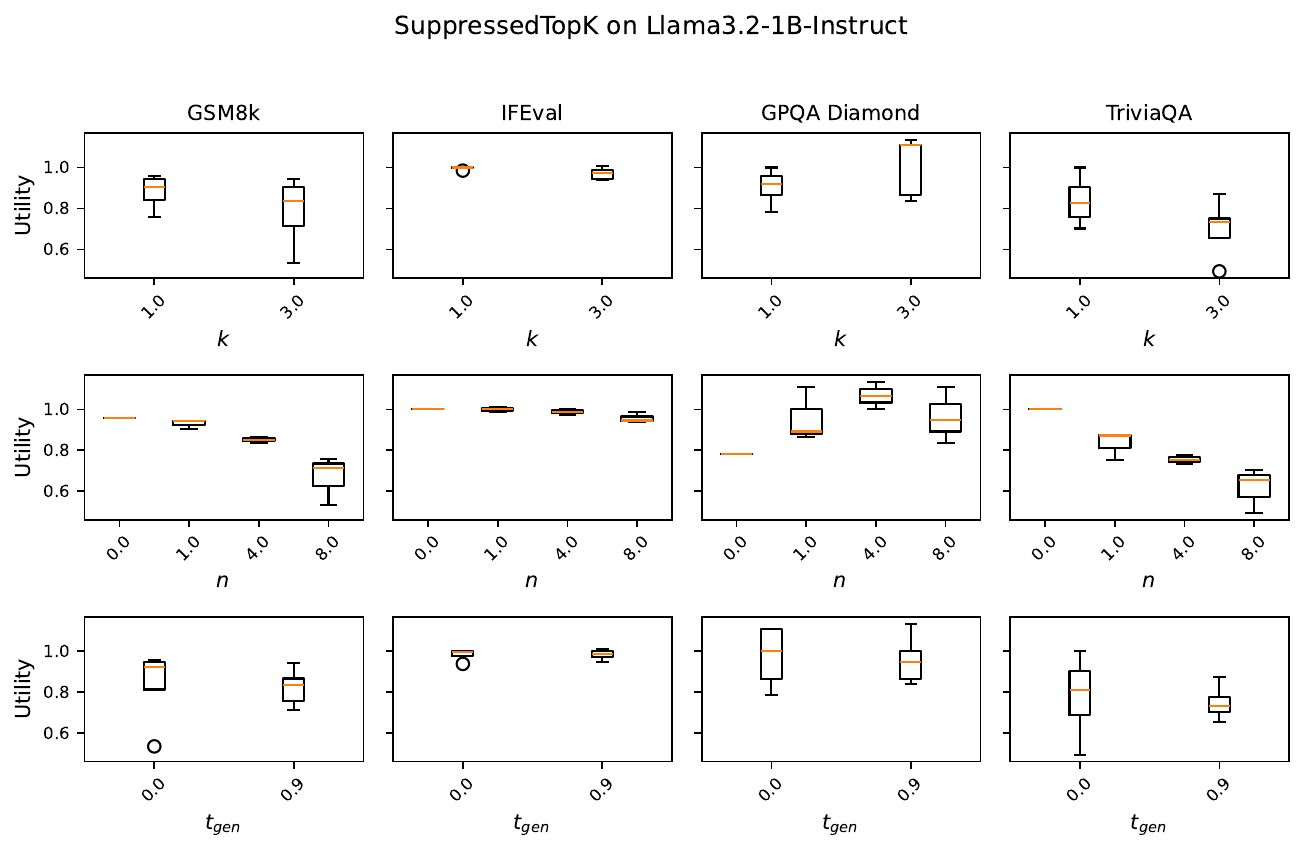}
   
    \includegraphics[width=0.9\linewidth]{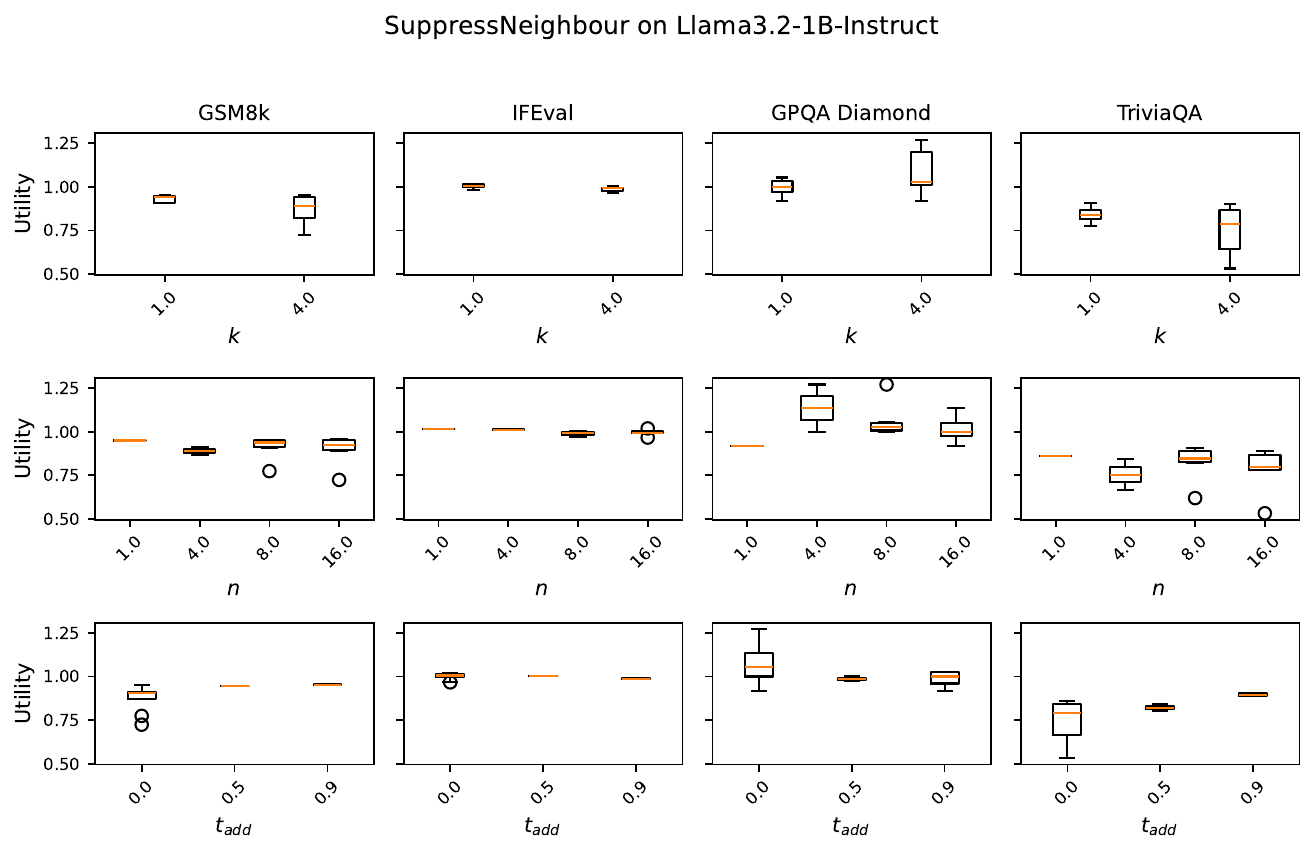}
     \caption{Detailed utility results for Llama-3.2-1B-Instruct under attack}
    \label{fig:detailed-attacks-util-llama-3-1b}
\end{figure*}

\begin{figure*}
    \centering
    \includegraphics[width=0.9\linewidth]{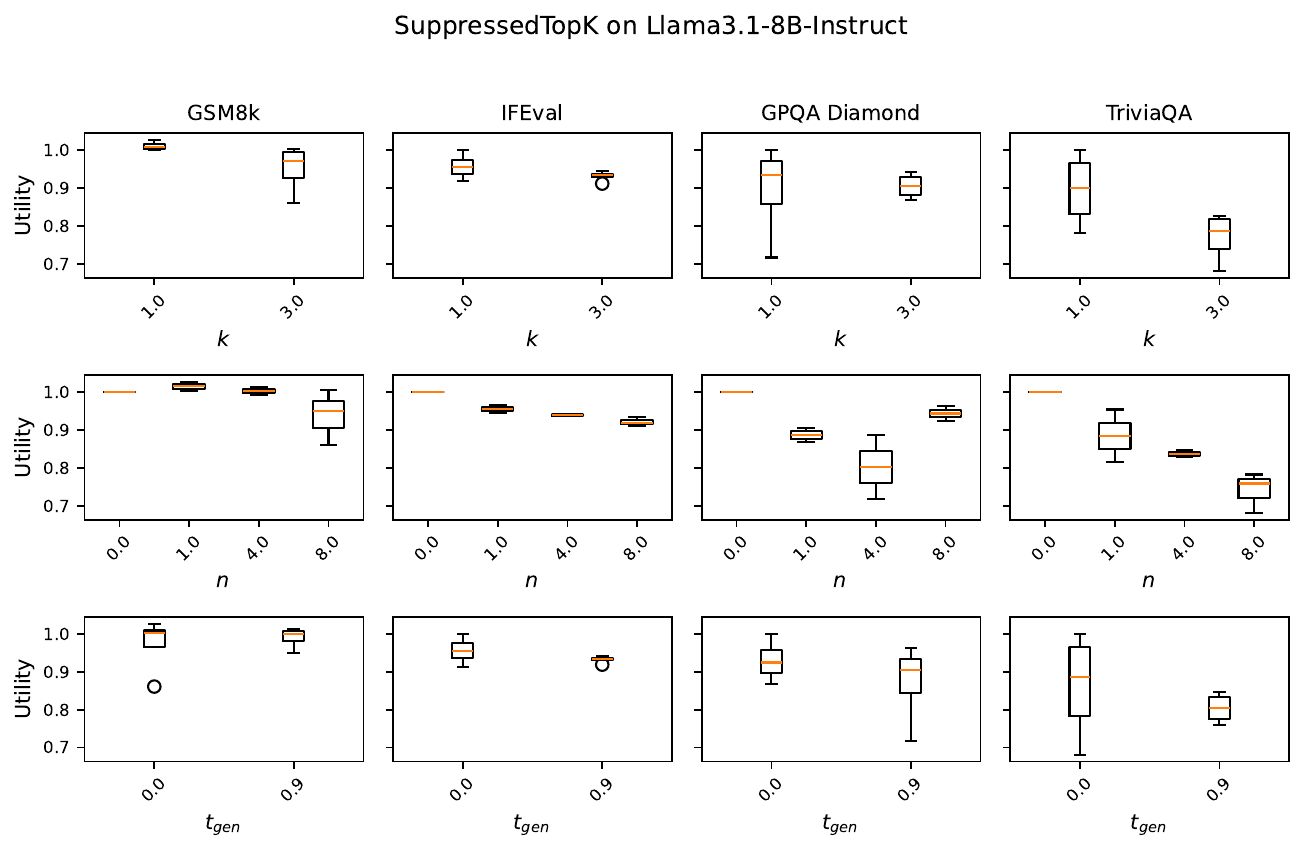}

   \includegraphics[width=0.9\linewidth]{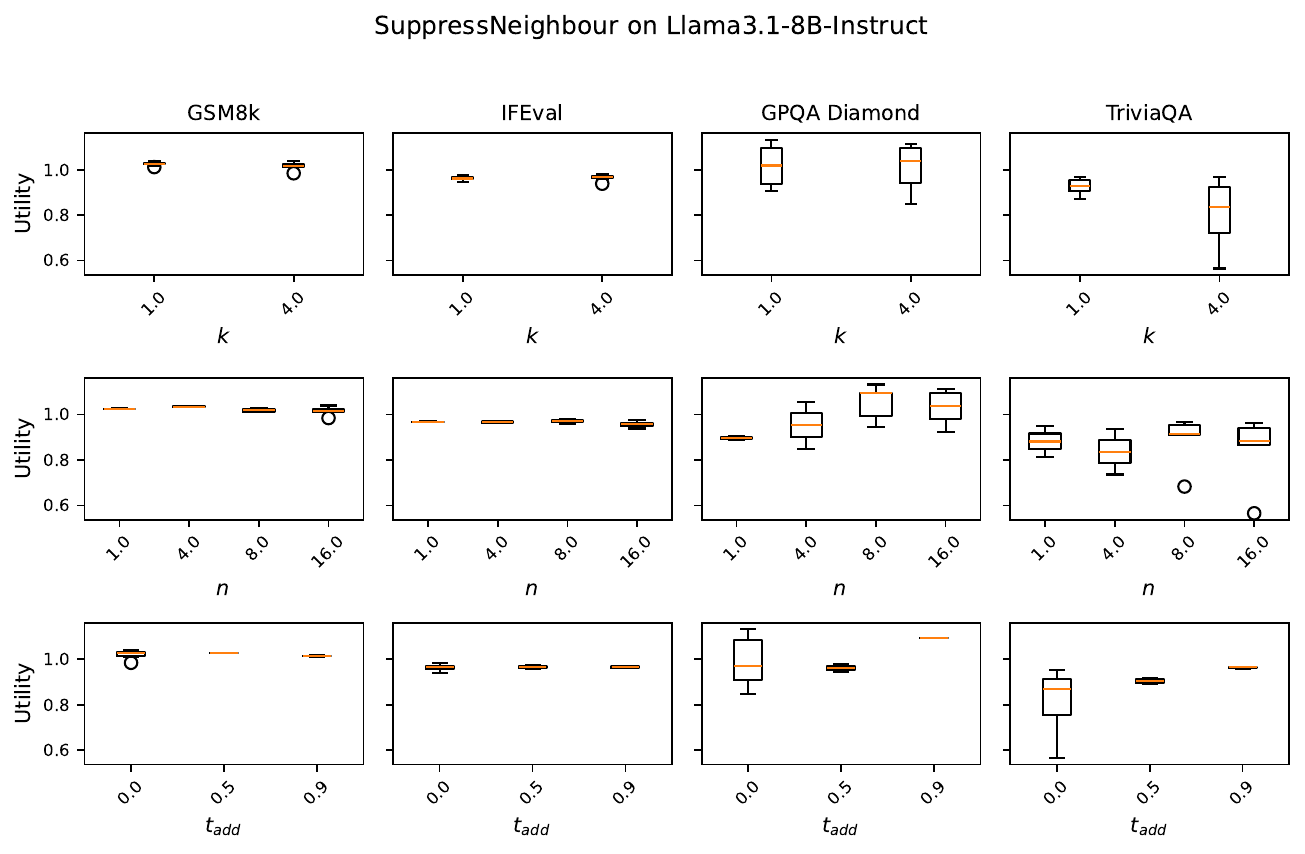}
     \caption{Detailed utility results for Llama-3.1-8B-Instruct under attack}
    \label{fig:detailed-attacks-util-llama-3-8b}
\end{figure*}

We plot the detailed dataset wise utility results for Llama 3.2-1B-Instruct ~\cref{fig:detailed-attacks-util-llama-3-1b} and Llama-3.1-8B-Instruct~\cref{fig:detailed-attacks-util-llama-3-8b}. We find that \btwattacknk is more robust to hyperparameters like number of attack steps $n$ as compared to \itattacknosp. We also see that TriviaQA is the most sensitive to number of attack steps, since it expects short factual answers. Finally, we find that larger models are more robust to utility drops induced by sampling 

\subsection{Detailed results by fingerprinting scheme}
\label{app:more-results}
\begin{figure*}[t]
    \centering
    \includegraphics[width=0.4\textwidth]{figs/detailed/legend_instructional_fp.pdf}
    \vspace{0.4em}

    \begin{subfigure}[b]{0.24\textwidth}
        \centering
        \includegraphics[width=\linewidth]{figs/detailed/asr_vs_util_llama-3.2-1b-instruct_16_instructional_fp.pdf}
        \caption{Llama-3.2-1B-Instruct (16)}
    \end{subfigure}\hfill
    \begin{subfigure}[b]{0.24\textwidth}
        \centering
        \includegraphics[width=\linewidth]{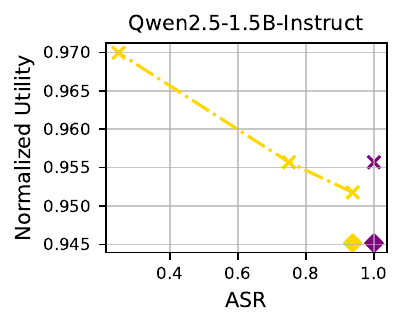}
        \caption{Qwen2.5-1.5B-Instruct (16)}
    \end{subfigure}\hfill
    \begin{subfigure}[b]{0.24\textwidth}
        \centering
        \includegraphics[width=\linewidth]{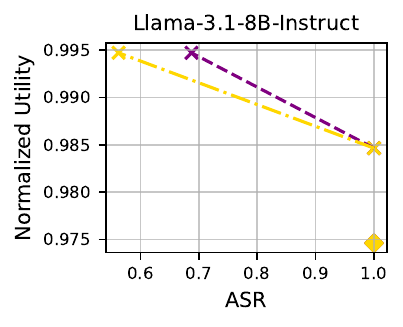}
        \caption{Llama-3.1-8B-Instruct (16)}
    \end{subfigure}\hfill
    \begin{subfigure}[b]{0.24\textwidth}
        \centering
        \includegraphics[width=\linewidth]{figs/detailed/asr_vs_util_qwen2.5-7b-instruct_16_instructional_fp.pdf}
        \caption{Qwen2.5-7B-Instruct (16)}
    \end{subfigure}

    \begin{subfigure}[b]{0.24\textwidth}
        \centering
        \includegraphics[width=\linewidth]{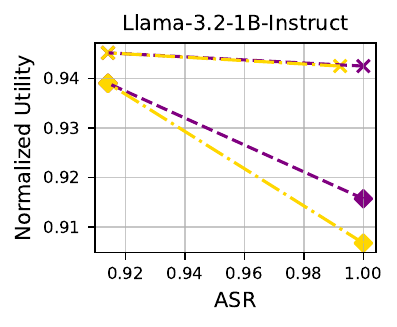}
        \caption{Llama-3.2-1B-Instruct (128)}
    \end{subfigure}
    \begin{subfigure}[b]{0.24\textwidth}
        \centering
        \includegraphics[width=\linewidth]{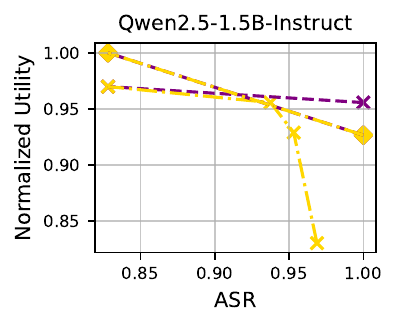}
        \caption{Qwen2.5-1.5B-Instruct (128)}
    \end{subfigure}

    \caption{Detailed ASR--Utility trade-offs for \textbf{Instructional Fingerprints}~\cite{xu2024instructionalfingerprintinglargelanguage}. Top row: 16 fingerprints; bottom row: 128 fingerprints.}
    \label{fig:app-detailed-instructional}
\end{figure*}

\begin{figure*}[t]
    \centering
    \includegraphics[width=0.4\textwidth]{figs/detailed/legend_chain_hash.pdf}
    \vspace{0.4em}

    \begin{subfigure}[b]{0.24\textwidth}
        \centering
        \includegraphics[width=\linewidth]{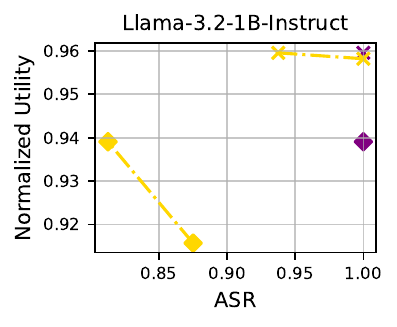}
        \caption{Llama-3.2-1B-Instruct (16)}
    \end{subfigure}\hfill
    \begin{subfigure}[b]{0.24\textwidth}
        \centering
        \includegraphics[width=\linewidth]{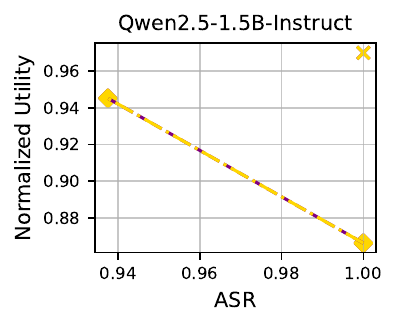}
        \caption{Qwen2.5-1.5B-Instruct (16)}
    \end{subfigure}\hfill
    \begin{subfigure}[b]{0.24\textwidth}
        \centering
        \includegraphics[width=\linewidth]{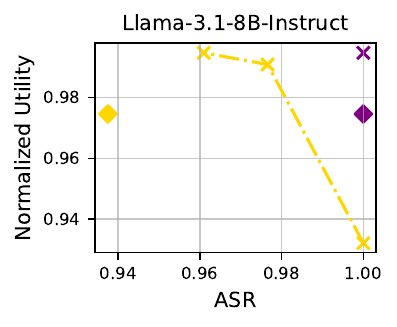}
        \caption{Llama-3.1-8B-Instruct (16)}
    \end{subfigure}\hfill
    \begin{subfigure}[b]{0.24\textwidth}
        \centering
        \includegraphics[width=\linewidth]{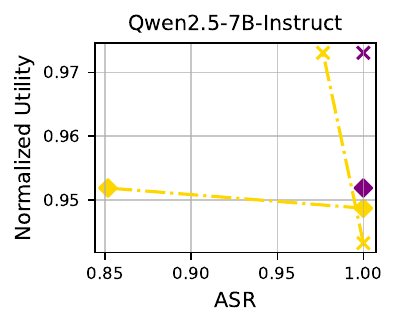}
        \caption{Qwen2.5-7B-Instruct (16)}
    \end{subfigure}

    \begin{subfigure}[b]{0.24\textwidth}
        \centering
        \includegraphics[width=\linewidth]{figs/detailed/asr_vs_util_llama-3.2-1b-instruct_128_chain_hash.pdf}
        \caption{Llama-3.2-1B-Instruct (128)}
    \end{subfigure}\hfill
    \begin{subfigure}[b]{0.24\textwidth}
        \centering
        \includegraphics[width=\linewidth]{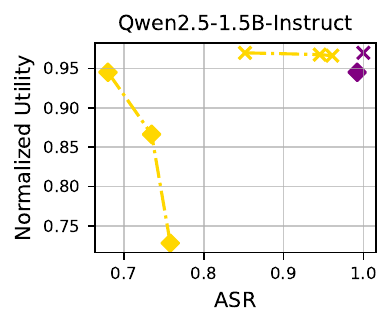}
        \caption{Qwen2.5-1.5B-Instruct (128)}
    \end{subfigure}\hfill
    \begin{subfigure}[b]{0.24\textwidth}
        \centering
        \includegraphics[width=\linewidth]{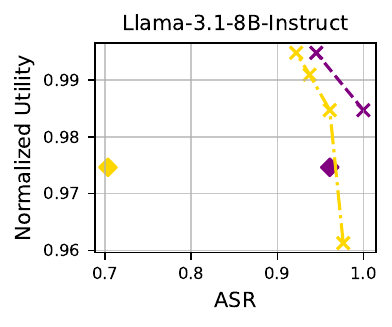}
        \caption{Llama-3.1-8B-Instruct (128)}
    \end{subfigure}\hfill
    \begin{subfigure}[b]{0.24\textwidth}
        \centering
        \includegraphics[width=\linewidth]{figs/detailed/asr_vs_util_qwen2.5-7b-instruct_128_chain_hash.pdf}
        \caption{Qwen2.5-7B-Instruct (128)}
    \end{subfigure}

    \caption{Detailed ASR--Utility trade-offs for \textbf{Chain\&Hash}~\cite{russinovich2024heythatsmodelintroducing}. Top row: 16 fingerprints; bottom row: 128 fingerprints.}
    \label{fig:app-detailed-chainhash}
\end{figure*}

\begin{figure*}[t]
    \centering
    \includegraphics[width=0.4\textwidth]{figs/detailed/legend_perinucleus.pdf}
    \vspace{0.4em}

    \begin{subfigure}[b]{0.24\textwidth}
        \centering
        \includegraphics[width=\linewidth]{figs/detailed/asr_vs_util_llama-3.2-1b-instruct_16_perinucleus.pdf}
        \caption{Llama-3.2-1B-Instruct (16)}
    \end{subfigure}\hfill
    \begin{subfigure}[b]{0.24\textwidth}
        \centering
        \includegraphics[width=\linewidth]{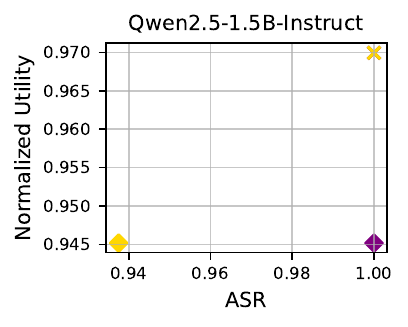}
        \caption{Qwen2.5-1.5B-Instruct (16)}
    \end{subfigure}\hfill
    \begin{subfigure}[b]{0.24\textwidth}
        \centering
        \includegraphics[width=\linewidth]{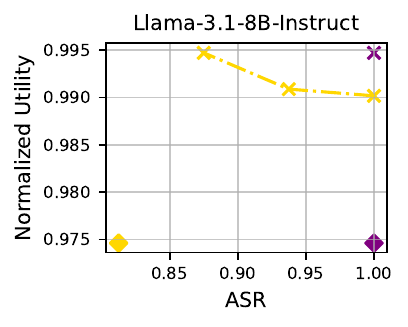}
        \caption{Llama-3.1-8B-Instruct (16)}
    \end{subfigure}\hfill
    \begin{subfigure}[b]{0.24\textwidth}
        \centering
        \includegraphics[width=\linewidth]{figs/detailed/asr_vs_util_qwen2.5-7b-instruct_16_perinucleus.pdf}
        \caption{Qwen2.5-7B-Instruct (16)}
    \end{subfigure}

    \begin{subfigure}[b]{0.24\textwidth}
        \centering
        \includegraphics[width=\linewidth]{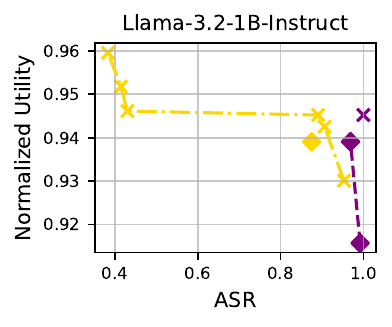}
        \caption{Llama-3.2-1B-Instruct (128)}
    \end{subfigure}\hfill
    \begin{subfigure}[b]{0.24\textwidth}
        \centering
        \includegraphics[width=\linewidth]{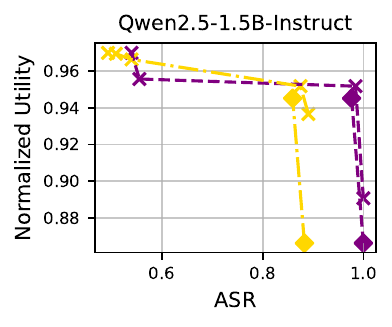}
        \caption{Qwen2.5-1.5B-Instruct (128)}
    \end{subfigure}\hfill
    \begin{subfigure}[b]{0.24\textwidth}
        \centering
        \includegraphics[width=\linewidth]{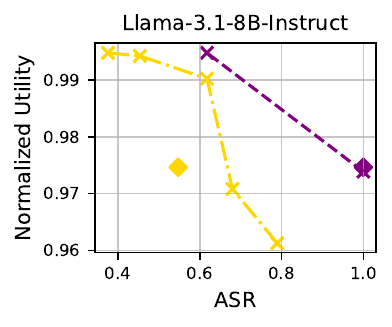}
        \caption{Llama-3.1-8B-Instruct (128)}
    \end{subfigure}\hfill
    \begin{subfigure}[b]{0.24\textwidth}
        \centering
        \includegraphics[width=\linewidth]{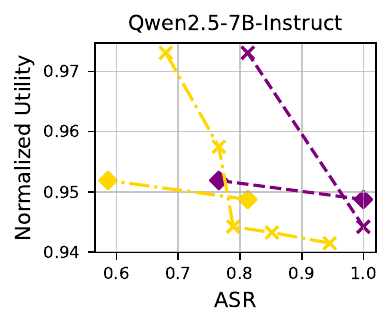}
        \caption{Qwen2.5-7B-Instruct (128)}
    \end{subfigure}

    \caption{Detailed ASR--Utility trade-offs for \textbf{Perinucleus Fingerprints}~\cite{nasery2025scalable}. Top row: 16 fingerprints; bottom row: 128 fingerprints.}
    \label{fig:app-detailed-perinucleus}
\end{figure*}

\begin{figure*}[t]
    \centering
    \includegraphics[width=0.4\textwidth]{figs/detailed/legend_imf.pdf}
    \vspace{0.4em}

    \begin{subfigure}[b]{0.24\textwidth}
        \centering
        \includegraphics[width=\linewidth]{figs/detailed/asr_vs_util_llama-3.2-1b-instruct_16_imf.pdf}
        \caption{Llama-3.2-1B-Instruct (16)}
    \end{subfigure}\hfill
    \begin{subfigure}[b]{0.24\textwidth}
        \centering
        \includegraphics[width=\linewidth]{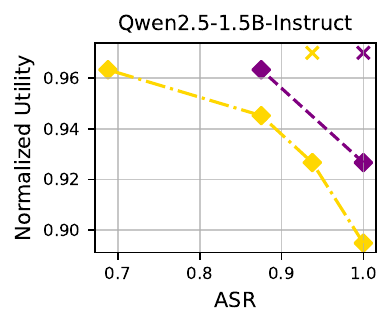}
        \caption{Qwen2.5-1.5B-Instruct (16)}
    \end{subfigure}\hfill
    \begin{subfigure}[b]{0.24\textwidth}
        \centering
        \includegraphics[width=\linewidth]{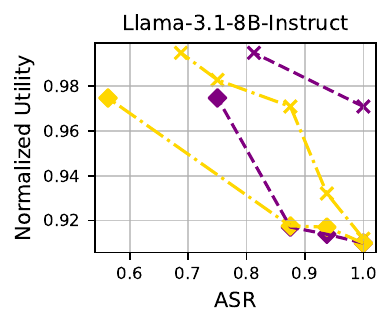}
        \caption{Llama-3.1-8B-Instruct (16)}
    \end{subfigure}\hfill
    \begin{subfigure}[b]{0.24\textwidth}
        \centering
        \includegraphics[width=\linewidth]{figs/detailed/asr_vs_util_qwen2.5-7b-instruct_16_imf.pdf}
        \caption{Qwen2.5-7B-Instruct (16)}
    \end{subfigure}

    \begin{subfigure}[b]{0.24\textwidth}
        \centering
        \includegraphics[width=\linewidth]{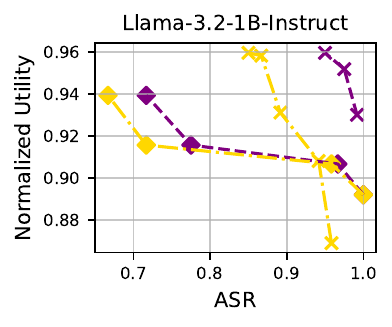}
        \caption{Llama-3.2-1B-Instruct (128)}
    \end{subfigure}\hfill
    \begin{subfigure}[b]{0.24\textwidth}
        \centering
        \includegraphics[width=\linewidth]{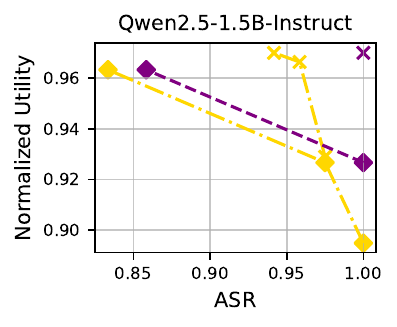}
        \caption{Qwen2.5-1.5B-Instruct (128)}
    \end{subfigure}\hfill
    \begin{subfigure}[b]{0.24\textwidth}
        \centering
        \includegraphics[width=\linewidth]{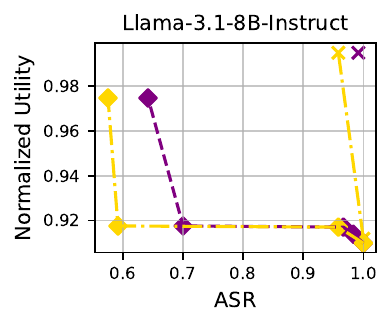}
        \caption{Llama-3.1-8B-Instruct (128)}
    \end{subfigure}\hfill
    \begin{subfigure}[b]{0.24\textwidth}
        \centering
        \includegraphics[width=\linewidth]{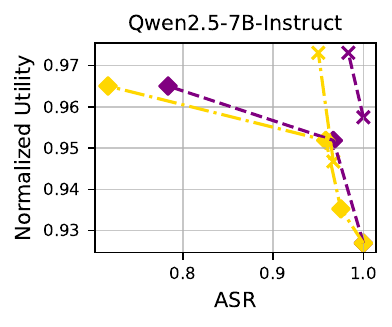}
        \caption{Qwen2.5-7B-Instruct (128)}
    \end{subfigure}

    \caption{Detailed ASR--Utility trade-offs for \textbf{Implicit Fingerprints (ImF)}~\cite{jiaxuan2025imfimplicitfingerprintlarge}. Top row: 16 fingerprints; bottom row: 128 fingerprints.}
    \label{fig:app-detailed-imf}
\end{figure*}

\begin{figure*}[t]
    \centering
    \includegraphics[width=0.4\textwidth]{figs/detailed/legend_imf.pdf}
    \vspace{0.4em}

    \begin{subfigure}[b]{0.24\textwidth}
        \centering
        \includegraphics[width=\linewidth]{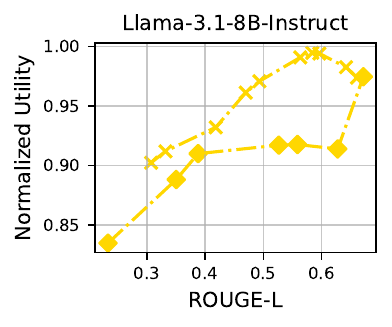}
        \caption{Llama-3.2-1B (16)}
    \end{subfigure}\hfill
    \begin{subfigure}[b]{0.24\textwidth}
        \centering
        \includegraphics[width=\linewidth]{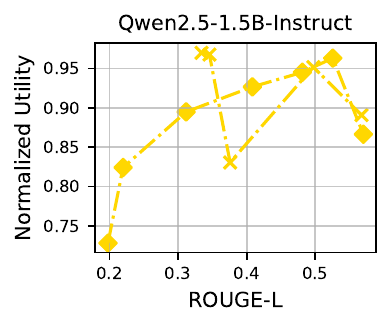}
        \caption{Qwen2.5-1.5B-Instruct (16)}
    \end{subfigure}\hfill
    \begin{subfigure}[b]{0.24\textwidth}
        \centering
        \includegraphics[width=\linewidth]{figs/detailed/rouge_l_vs_util_llama-3.1-8b-instruct_16_imf.pdf}
        \caption{Llama-3.1-8B-Instruct (16)}
    \end{subfigure}\hfill
    \begin{subfigure}[b]{0.24\textwidth}
        \centering
        \includegraphics[width=\linewidth]{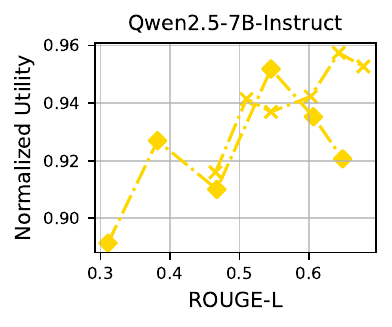}
        \caption{Qwen2.5-7B-Instruct (16)}
    \end{subfigure}

    \begin{subfigure}[b]{0.24\textwidth}
        \centering
        \includegraphics[width=\linewidth]{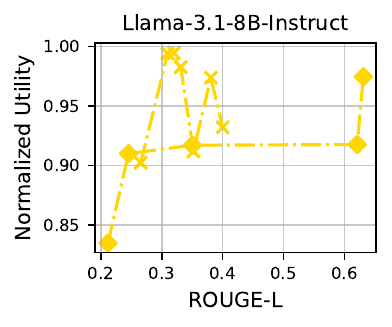}
        \caption{Llama-3.2-1B-Instruct (128)}
    \end{subfigure}\hfill
    \begin{subfigure}[b]{0.24\textwidth}
        \centering
        \includegraphics[width=\linewidth]{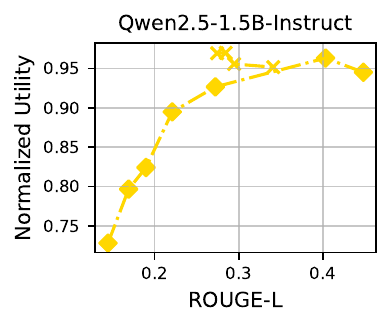}
        \caption{Qwen2.5-1.5B-Instruct (128)}
    \end{subfigure}\hfill
    \begin{subfigure}[b]{0.24\textwidth}
        \centering
        \includegraphics[width=\linewidth]{figs/detailed/rouge_l_vs_util_llama-3.1-8b-instruct_128_imf.pdf}
        \caption{Llama-3.1-8B-Instruct (128)}
    \end{subfigure}\hfill
    \begin{subfigure}[b]{0.24\textwidth}
        \centering
        \includegraphics[width=\linewidth]{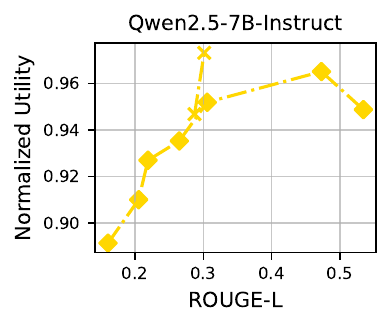}
        \caption{Qwen2.5-7B-Instruct (128)}
    \end{subfigure}

    \caption{Detailed ROUGE-L--Utility trade-offs for \textbf{Implicit Fingerprints (ImF)}~\cite{jiaxuan2025imfimplicitfingerprintlarge}. Top row: 16 fingerprints; bottom row: 128 fingerprints.}
    \label{fig:app-detailed-imf-rouge}
\end{figure*}

\begin{figure*}[t]
    \centering
    \includegraphics[width=0.4\textwidth]{figs/detailed/legend_fpedit.pdf}
    \vspace{0.4em}

    \begin{subfigure}[b]{0.24\textwidth}
        \centering
        \includegraphics[width=\linewidth]{figs/detailed/asr_vs_util_llama-3.2-1b-instruct_16_fpedit.pdf}
        \caption{Llama-3.2-1B-Instruct (16)}
    \end{subfigure}\hfill
    \begin{subfigure}[b]{0.24\textwidth}
        \centering
        \includegraphics[width=\linewidth]{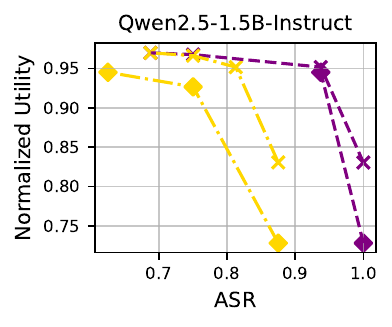}
        \caption{Qwen2.5-1.5B-Instruct (16)}
    \end{subfigure}\hfill
    \begin{subfigure}[b]{0.24\textwidth}
        \centering
        \includegraphics[width=\linewidth]{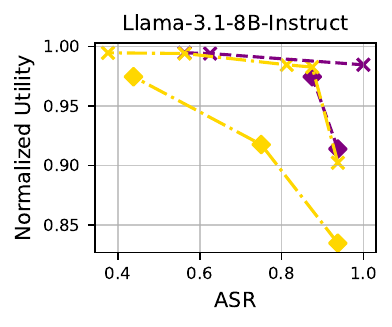}
        \caption{Llama-3.1-8B-Instruct (16)}
    \end{subfigure}\hfill
    \begin{subfigure}[b]{0.24\textwidth}
        \centering
        \includegraphics[width=\linewidth]{figs/detailed/asr_vs_util_qwen2.5-7b-instruct_16_fpedit.pdf}
        \caption{Qwen2.5-7B-Instruct (16)}
    \end{subfigure}

    \begin{subfigure}[b]{0.24\textwidth}
        \centering
        \includegraphics[width=\linewidth]{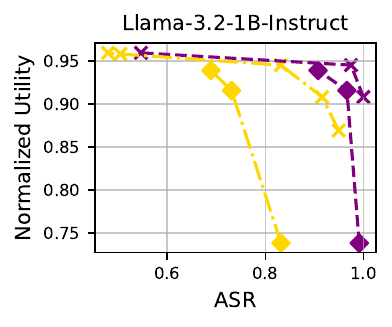}
        \caption{Llama-3.2-1B-Instruct (128)}
    \end{subfigure}\hfill
    \begin{subfigure}[b]{0.24\textwidth}
        \centering
        \includegraphics[width=\linewidth]{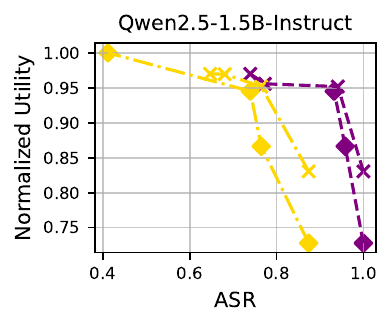}
        \caption{Qwen2.5-1.5B-Instruct (128)}
    \end{subfigure}\hfill
    \begin{subfigure}[b]{0.24\textwidth}
        \centering
        \includegraphics[width=\linewidth]{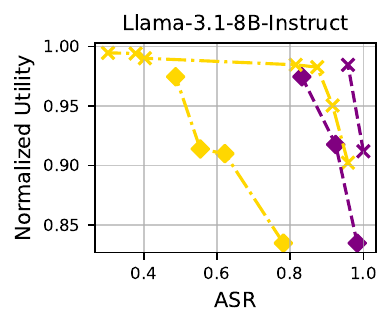}
        \caption{Llama-3.1-8B-Instruct (128)}
    \end{subfigure}\hfill
    \begin{subfigure}[b]{0.24\textwidth}
        \centering
        \includegraphics[width=\linewidth]{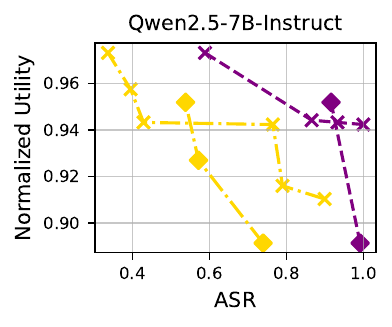}
        \caption{Qwen2.5-7B-Instruct (128)}
    \end{subfigure}

    \caption{Detailed ASR--Utility trade-offs for \textbf{FPEdit}~\cite{wang2025fpeditrobustllmfingerprinting}. Top row: 16 fingerprints; bottom row: 128 fingerprints.}
    \label{fig:app-detailed-fpedit}
\end{figure*}

\begin{figure*}[t]
    \centering
    \begin{subfigure}[b]{0.24\textwidth}
        \centering
        \includegraphics[width=\linewidth]{figs/detailed/dsw_util_cherry_picked_meta-llama_llama-3.2-1b_prim.pdf}
        \caption{Llama-3.2-1B-Instruct}
    \end{subfigure}\hfill
    \begin{subfigure}[b]{0.24\textwidth}
        \centering
        \includegraphics[width=\linewidth]{figs/detailed/dsw_util_cherry_picked_meta-llama_llama-3.2-1b_sec.pdf}
        \caption{Llama-3.2-1B-Instruct}
    \end{subfigure}\hfill
    \begin{subfigure}[b]{0.24\textwidth}
        \centering
        \includegraphics[width=\linewidth]{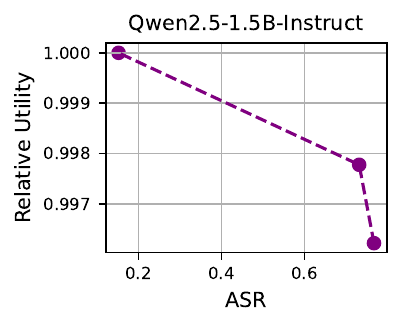}
        \caption{Qwen2.5-1.5B-Instruct}
    \end{subfigure}\hfill
    \begin{subfigure}[b]{0.24\textwidth}
        \centering
        \includegraphics[width=\linewidth]{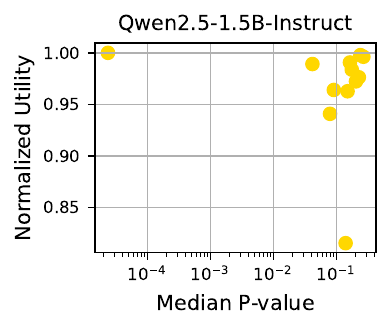}
        \caption{Qwen2.5-1.5B-Instruct}
    \end{subfigure}

    \caption{Detailed ASR--Utility trade-offs for \textbf{Domain Specific Watermarks}~\cite{gloaguen2025robustllmfingerprintingdomainspecific}.}
    \label{fig:app-detailed-dsw}
\end{figure*}

\begin{figure*}[t]
    \centering
    \includegraphics[width=0.3\textwidth]{figs/legend_pplx.pdf}
    \vspace{0.4em}
    
    \begin{subfigure}[b]{0.24\textwidth}
        \centering
        \includegraphics[width=\linewidth]{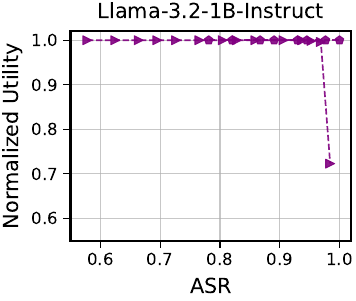}
        \caption{Llama-3.2-1B-Instruct (128)}
    \end{subfigure}\hfill
    \begin{subfigure}[b]{0.24\textwidth}
        \centering
        \includegraphics[width=\linewidth]{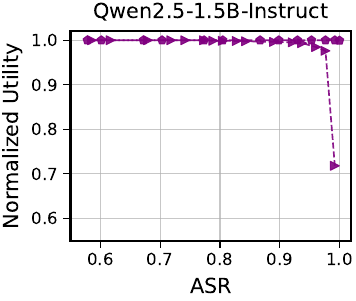}
        \caption{Qwen2.5-1.5B-Instruct (128)}
    \end{subfigure}\hfill
    \begin{subfigure}[b]{0.24\textwidth}
        \centering
        \includegraphics[width=\linewidth]{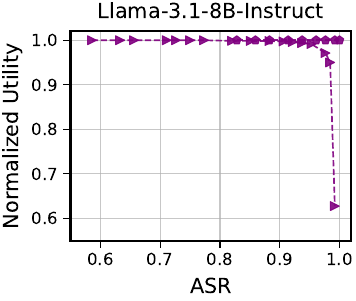}
        \caption{Llama-3.1-8B-Instruct (128)}
    \end{subfigure}\hfill
    \begin{subfigure}[b]{0.24\textwidth}
        \centering
        \includegraphics[width=\linewidth]{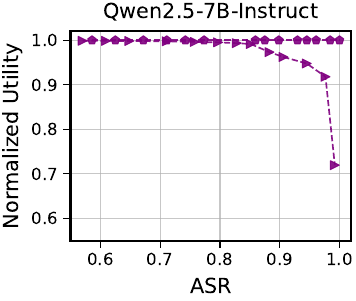}
        \caption{Qwen2.5-7B-Instruct (128)}
    \end{subfigure}

    \caption{Detailed ASR--Utility trade-offs for 128 \textbf{RoFL Fingerprints}~\cite{tsai2025rofl}.}
    \label{fig:app-detailed-rofl}
\end{figure*}

\begin{figure*}[t]
    \centering
    \includegraphics[width=0.3\textwidth]{figs/legend_pplx.pdf}
    \vspace{0.4em}
    
    \begin{subfigure}[b]{0.24\textwidth}
        \centering
        \includegraphics[width=\linewidth]{figs/detailed_multirofl/final_asr_util_llama32_1b_roflmulti_paper_like.pdf}
        \caption{Llama-3.2-1B-Instruct (128)}
    \end{subfigure}\hfill
    \begin{subfigure}[b]{0.24\textwidth}
        \centering
        \includegraphics[width=\linewidth]{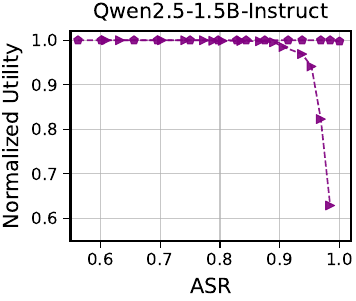}
        \caption{Qwen2.5-1.5B-Instruct (128)}
    \end{subfigure}\hfill
    \begin{subfigure}[b]{0.24\textwidth}
        \centering
        \includegraphics[width=\linewidth]{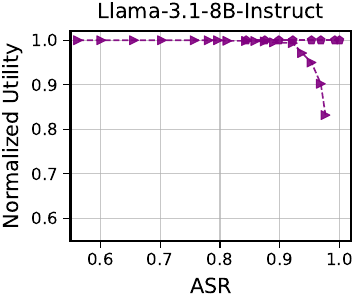}
        \caption{Llama-3.1-8B-Instruct (128)}
    \end{subfigure}\hfill
    \begin{subfigure}[b]{0.24\textwidth}
        \centering
        \includegraphics[width=\linewidth]{figs/detailed_multirofl/final_asr_util_qwen25_7b_roflmulti_paper_like.pdf}
        \caption{Qwen2.5-7B-Instruct (128)}
    \end{subfigure}

    \caption{Detailed ASR--Utility trade-offs for 128 \textbf{(Multi) RoFL Fingerprints}~\cite{tsai2025rofl}.}
    \label{fig:app-detailed-multi-rofl}
\end{figure*}

\begin{figure*}[t]
    \centering
    \includegraphics[width=0.3\textwidth]{figs/legend_pplx.pdf}
    \vspace{0.4em}
    
    \begin{subfigure}[b]{0.24\textwidth}
        \centering
        \includegraphics[width=\linewidth]{figs/detailed_proflingo/final_asr_util_llama32_1b_proflingo_paper_like.pdf}
        \caption{Llama-3.2-1B-Instruct (128)}
    \end{subfigure}\hfill
    \begin{subfigure}[b]{0.24\textwidth}
        \centering
        \includegraphics[width=\linewidth]{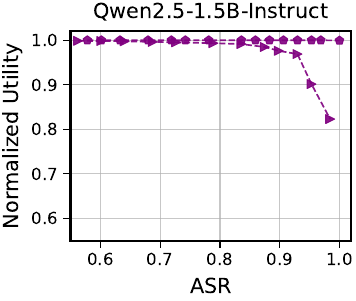}
        \caption{Qwen2.5-1.5B-Instruct (128)}
    \end{subfigure}\hfill
    \begin{subfigure}[b]{0.24\textwidth}
        \centering
        \includegraphics[width=\linewidth]{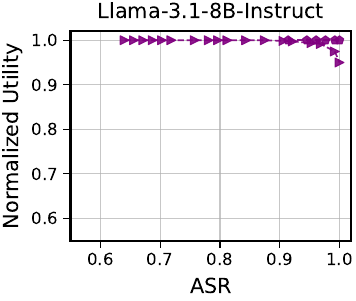}
        \caption{Llama-3.1-8B-Instruct (128)}
    \end{subfigure}\hfill
    \begin{subfigure}[b]{0.24\textwidth}
        \centering
        \includegraphics[width=\linewidth]{figs/detailed_proflingo/final_asr_util_qwen25_7b_proflingo_paper_like.pdf}
        \caption{Qwen2.5-7B-Instruct (128)}
    \end{subfigure}

    \caption{Detailed ASR--Utility trade-offs for 128 \textbf{ProfLingo Fingerprints}~\cite{jin2024proflingo}.}
    \label{fig:app-detailed-proflingo}
\end{figure*}

\begin{figure*}[t]
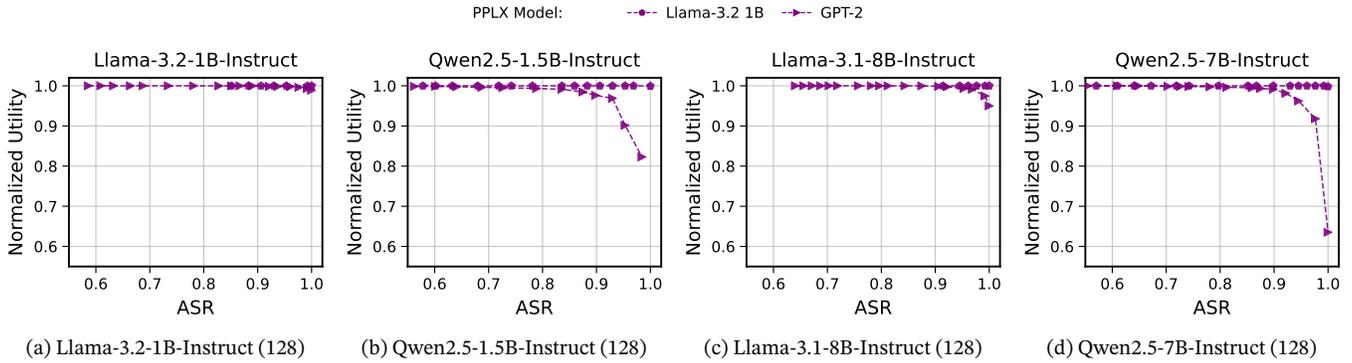

    \centering
    \includegraphics[width=0.3\textwidth]{figs/legend_pplx.pdf}
    \vspace{0.4em}

    \begin{subfigure}[b]{0.24\textwidth}
        \centering
        \includegraphics[width=\linewidth]{figs/detailed_proflingo/final_asr_util_llama32_1b_proflingo_paper_like.pdf}
        \caption{Llama-3.2-1B-Instruct (128)}
    \end{subfigure}\hfill
    \begin{subfigure}[b]{0.24\textwidth}
        \centering
        \includegraphics[width=\linewidth]{figs/detailed_proflingo/final_asr_util_qwen25_15b_proflingo_paper_like.pdf}
        \caption{Qwen2.5-1.5B-Instruct (128)}
    \end{subfigure}\hfill
    \begin{subfigure}[b]{0.24\textwidth}
        \centering
        \includegraphics[width=\linewidth]{figs/detailed_proflingo/final_asr_util_llama31_8b_proflingo_paper_like.pdf}
        \caption{Llama-3.1-8B-Instruct (128)}
    \end{subfigure}\hfill
    \begin{subfigure}[b]{0.24\textwidth}
        \centering
        \includegraphics[width=\linewidth]{figs/detailed_proflingo/final_asr_util_qwen25_7b_proflingo_paper_like.pdf}
        \caption{Qwen2.5-7B-Instruct (128)}
    \end{subfigure}

    \caption{Detailed ASR--Utility trade-offs for 128 \textbf{MergePrint Fingerprints}~\cite{jin2024proflingo}.}
    \label{fig:app-detailed-mergeprint}
\end{figure*}

We provide detailed ASR--Utility curves across four models and upto two fingerprint-set sizes for each fingerprinting scheme in \cref{fig:app-detailed-instructional}-\cref{fig:app-detailed-mergeprint}. 

\subsection{Other experimental details}
For RoFL~\cite{tsai2025rofl} we use the models in Figure \ref{fig:shadow_models} as shadow models to optimize the fingerprints. 

For DomainSpecificWatermarks~\cite{gloaguen2025robustllmfingerprintingdomainspecific}, we pick the 6000 most common English words for computing the green list. We also generate texts using these words for statistical analysis. We clip the stats at 99.9th percentile for each token to be robust to extreme outliers. We experiment with sparsity using top-p and top-k on $p_{model}(\cdot|t_i)$. We use a base model (e.g. Llama-3.2-1B-Base, Llama-3.2-3B-Base) as $p_{base}$ for calibration. During detection we use the prompts from~\cite{gloaguen2025robustllmfingerprintingdomainspecific} for the medical domain. We vary $\delta$ in \{2,4,8,12\} for scrubbing the watermark.

\begin{figure*}[h!]
\centering
\resizebox{\textwidth}{!}{%
\begin{tabular}{|l|l|l|}
\hline
\textbf{Fingerprinted Model} & \textbf{Shadow Model 1} & \textbf{Shadow Model 2} \\ \hline
meta-llama/Llama-3.1-8B-Instruct & OpenLLM-Ro/RoLlama3.1-8b-Instruct & TsinghuaC3I/Llama-3.1-8B-UltraMedical \\ \hline
meta-llama/Llama-3.2-1B-Instruct & Vikhrmodels/Vikhr-Llama-3.2-1B-Instruct & phamhai/Llama-3.2-1B-Instruct-Frog \\ \hline
Qwen/Qwen2.5-7B-Instruct & Unbabel/M-Prometheus-7B & AQuarterMile/WritingBench-Critic-Model-Qwen-7B \\ \hline
Qwen/Qwen2.5-1.5B-Instruct & Xtra-Computing/XtraGPT-1.5B & katanemo/Arch-Router-1.5B \\ \hline
\end{tabular}}
\caption{Fingerprinted models and their respective shadow models for RoFL.}
\label{fig:shadow_models}
\end{figure*}

\end{document}